\documentclass[12pt]{article}
\usepackage{geometry}
\usepackage{graphicx} 
\graphicspath{{./plots/}}
\usepackage{amsmath}
\usepackage{amssymb}
\usepackage{siunitx}
\DeclareSIUnit{\kton}{kton}
\DeclareSIUnit{\yr}{yr}
\usepackage{authblk}
\usepackage{cite}
\usepackage{hyperref}
\usepackage{multirow}
\usepackage{makecell}
\usepackage{booktabs}
\usepackage{caption}
\usepackage{subcaption}
\usepackage{lineno}

\title{Characterizing Single-Signal Events from Atmospheric-Neutrino Neutral-Current Interactions in Large Liquid Scintillator Detectors}

\author[a,b]{Zhenning Qu\thanks{quzn@ihep.ac.cn}}
\author[c]{Jie Cheng\thanks{chengjie@ncepu.edu.cn}}
\author[b]{Wan-lei Guo\thanks{guowl@ihep.ac.cn}}
\author[b]{Gaosong Li\thanks{ligs@ihep.ac.cn}}
\author[a,b]{Yu-Feng Li\thanks{liyufeng@ihep.ac.cn}}
\author[b]{Liang-jian Wen\thanks{wenlj@ihep.ac.cn}}
\affil[a]{School of Physical Sciences, University of Chinese Academy of Sciences, Beijing 100049, China}
\affil[b]{Institute of High Energy Physics, Chinese Academy of Sciences, Beijing 100049, China}
\affil[c]{School of Nuclear Science and Engineering, North China Electric Power University, Beijing 102206, China}
\date{}
\begin{document}

\maketitle
\begin{abstract}
Neutral-current interactions of atmospheric neutrinos in large liquid scintillator detectors offer a new opportunity to study single-signal events (hereafter singles), characterized by a prompt energy deposition on the MeV-to-GeV scale and no identified delayed signal. In this work, we systematically investigate the model dependence of atmospheric-neutrino singles due to the primary neutrino-nucleus interaction, residual-nucleus de-excitation, and secondary interactions in the scintillator. Our results show that the dominant model dependence originates from the primary neutrino-nucleus interaction, especially for neutral-current processes on carbon, whereas de-excitation is essential for the singles selection yet leads to relatively small spectral variations among realistic models. Secondary-interaction effects are also subdominant overall. We further present the predicted event rates and prompt-energy spectra for neutral-current singles, along with the charged-current contribution. Separately, we estimate the low-energy contribution from elastic scattering of sub-\SI{100}{\MeV} atmospheric neutrinos on free protons. These results highlight the physics potential of current and future large liquid scintillator detectors, such as the Jiangmen Underground Neutrino Observatory, to study atmospheric-neutrino singles, probe neutrino-nucleus interaction models, and improve background estimates for rare-event searches.

\end{abstract}

\clearpage
\section{Introduction}

Atmospheric neutrinos, produced in the decay chains of pions and kaons generated by cosmic-ray interactions in the Earth's atmosphere, constitute a unique natural neutrino source. The atmospheric-neutrino flux spans energies from tens of MeV to beyond the TeV scale and includes all three neutrino flavors and antineutrinos, with propagation baselines ranging from tens of kilometres to the Earth's diameter. This broad coverage makes atmospheric neutrinos a powerful laboratory for studying neutrino oscillations, matter effects, and the neutrino mass ordering. Atmospheric neutrinos provided the first compelling evidence for neutrino oscillations~\cite{Super-Kamiokande:1998kpq} and have since remained a cornerstone of neutrino-oscillation studies. At the same time, their interactions in matter provide a complementary probe of neutrino-nucleus physics, particularly the neutral-current (NC) sector, which is comparatively difficult to access with other neutrino sources. Large water Cherenkov detectors, most notably Super-Kamiokande (Super-K), have traditionally played a leading role in these studies. The recent transition of Super-K to gadolinium-loaded water further underscores the importance of neutron tagging for atmospheric-neutrino measurements and rare-event searches~\cite{Super-Kamiokande:2021the}. Nevertheless, even with gadolinium loading, neutron-tagging efficiency in water Cherenkov detectors remains below that achievable in large liquid scintillator (LS) detectors. Moreover, the low-energy hadrons produced in neutrino interactions are often below the Cherenkov threshold, limiting the information that can be reconstructed from their final states. These complementary considerations motivate atmospheric-neutrino measurements with large LS detectors, such as the Jiangmen Underground Neutrino Observatory (JUNO)~\cite{JUNO:2015zny}.

Large LS detectors represent an emerging frontier for atmospheric-neutrino studies. Unlike water Cherenkov detectors, LS detectors offer intrinsically high efficiency for detecting neutrons via $\gamma$-ray emission from capture processes. This capability, together with superior energy resolution and lower detection thresholds, allows efficient atmospheric-neutrino event detection and categorization according to their delayed signal topology: (1) singles with no delayed signal, (2) double-coincidence signals involving neutron capture, and (3) multi- or triple-coincidence signals that include additional delayed signatures. Additional signatures from long-lived residual-nucleus decays are searched for only in events with a tagged neutron capture and are therefore not part of the singles selection. While recent studies have characterized double-coincidence signals as backgrounds for the diffuse supernova neutrino background (DSNB)~\cite{Cheng:2020aaw,Cheng:2020oko} search, and the KamLAND Collaboration has used neutron multiplicity to probe the strange axial coupling constant~\cite{KamLAND:2022ptk}, the ``single-signal'' channel remains relatively unexplored~\cite{Chauhan:2021fzu}.

Singles in LS detectors open a largely unexplored window into low-energy neutrino interactions. At prompt energies below about \SI{100}{\MeV}, these events are expected to be dominated by hadronic products from NC interactions. Such low-energy hadrons often fall below the Cherenkov threshold and are therefore largely invisible in water Cherenkov detectors, whereas scintillation light allows LS detectors to detect their energy deposition. This gives LS experiments unique access to NC interactions in a regime where direct experimental constraints, particularly for interactions on carbon, remain limited.

Realizing this opportunity, however, requires a reliable understanding of how neutrino interactions are translated into observable scintillation signals. The predicted singles spectrum depends on several model-dependent ingredients, including the nuclear model, final-state interactions (FSI), residual-nucleus de-excitation, and secondary interactions (SI) in the scintillator. Quantifying these effects is becoming increasingly important as JUNO begins to collect large samples of atmospheric-neutrino singles. At the same time, the sensitivity of this sample to neutrino--nucleus interaction models provides an opportunity to constrain their associated uncertainties. The importance of this sample extends beyond neutrino-interaction physics: singles can mimic or contribute to backgrounds in a broad range of rare-event searches, including searches for solar ${}^{8}\mathrm{B}$ neutrinos~\cite{JUNO:2020hqc}, boosted dark matter and dark-matter annihilation into sterile neutrinos~\cite{Chauhan:2021fzu}, certain proton decay modes~\cite{JUNO:2015zny}, and neutrinoless double-beta decay~\cite{juno_0vbb}. Moreover, the sub-\SI{100}{\MeV} singles sample is enriched in single-proton recoils, providing a useful control sample for validating the detector's timing, pulse-shape, and quenching responses.

These considerations motivate a systematic assessment of the model dependence of atmospheric-neutrino singles. In this paper, we focus on the neutral-current component of atmospheric-neutrino singles in large LS detectors, while including the charged-current contribution for comparison. We find that NC interactions dominate the singles spectrum below a prompt energy of about \SI{100}{\MeV}, and that the spread among neutrino--nucleus interaction models constitutes the leading systematic uncertainty in this region. Separately, we estimate the low-energy contribution from elastic scattering of sub-\SI{100}{\MeV} atmospheric neutrinos on free protons.

The remainder of this paper is organized as follows. In Sec.~\ref{sec:int}, we define the single-signal topology and describe the event-rate prediction framework. In Sec.~\ref{sec:processes}, we examine the relevant physical processes, with particular emphasis on de-excitation and SI models not considered in previous studies such as Ref.~\cite{Chauhan:2021fzu}. In Sec.~\ref{sec:result}, we compare the model configurations and present the predicted event rates and prompt-energy spectra. The physics implications are discussed in Sec.~\ref{sec:discussion}, and a summary and outlook are provided in Sec.~\ref{sec:sum}.

\section{Atmospheric Neutrino Single-Signal}
\label{sec:int}

\begin{figure}[htbp]
    \centering
    \includegraphics[width=0.95\linewidth]{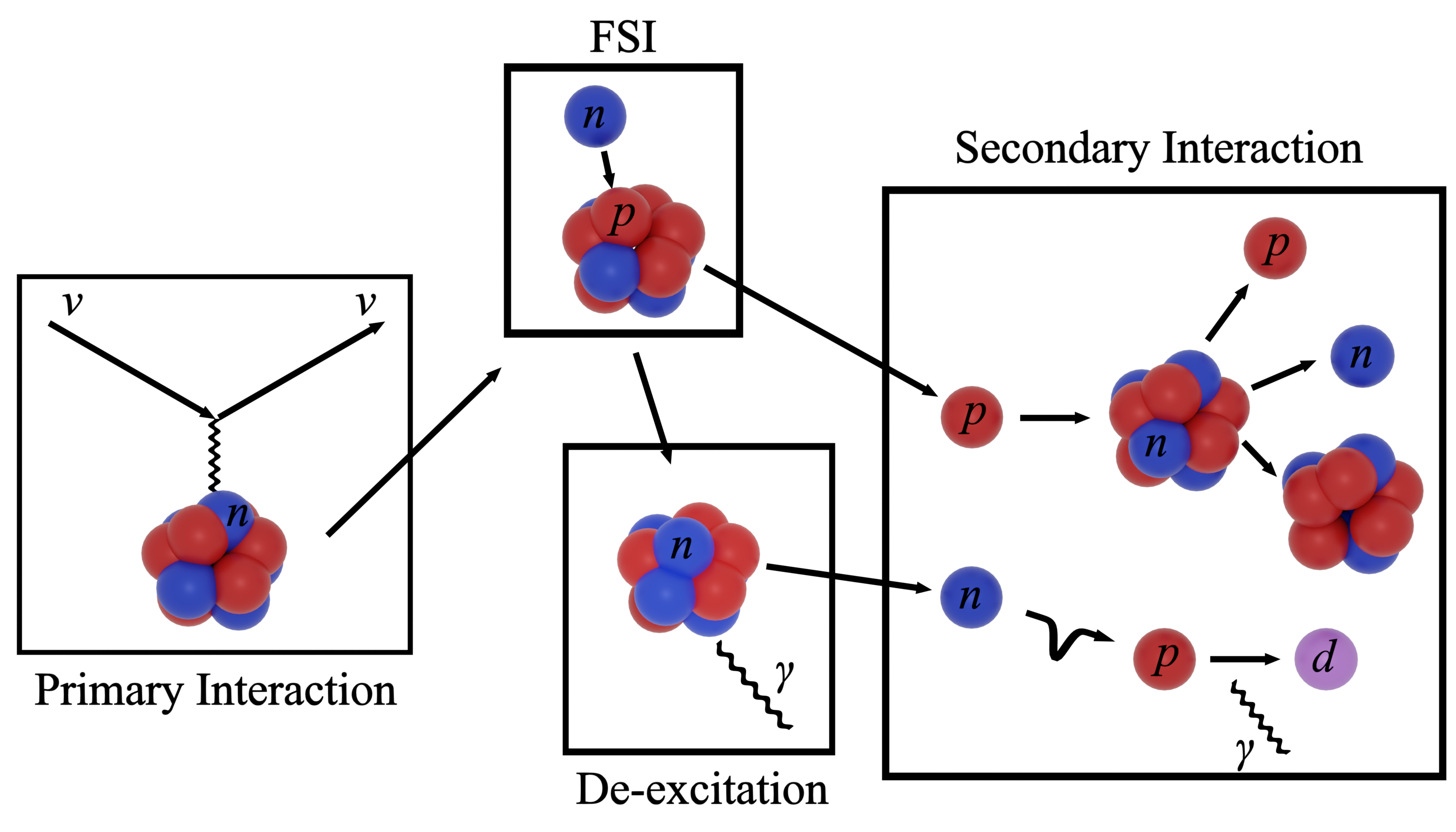}
    \caption{Schematic view of atmospheric-neutrino NC events in LS. The full chain includes the primary neutrino--nucleus interaction, intranuclear FSI, possible de-excitation of the residual nucleus, and secondary interactions of the outgoing particles in the LS.}
    \label{fig:schematic}
\end{figure}

The atmospheric-neutrino interaction in an LS detector proceeds through several distinct physical stages, as illustrated in Figure~\ref{fig:schematic}. The process begins with the primary interaction, in which a neutrino scatters off a target nucleus (typically ${}^{12}\mathrm{C}$ or ${}^{1}\mathrm{H}$) via either charged-current (CC) or NC exchange. Before escaping the nuclear environment, the resulting final-state particles may undergo FSI within the parent nucleus. Subsequently, the emitted particles propagate through the scintillator medium, depositing energy and potentially generating additional nucleons through SI. The residual nucleus, possibly left in an unstable or excited state, may de-excite until it reaches its ground or stable state. Neutrons produced throughout this chain eventually thermalize and are captured, predominantly by protons, yielding a characteristic \SI{2.223}{\MeV} $\gamma$ ray. Physically, the events of interest are those in which no neutrons are produced by the primary interaction. In the LS detector considered in this analysis, neutrons are tagged through their delayed neutron-capture signals, and the correspondence between a neutron-free primary final state and the observed delayed-signal topology is affected by FSI, de-excitation, and SI. In this study, we therefore refer to events with only a prompt energy deposition and no identified delayed signal as singles.

\subsection{Methodology for Single-Signal Event Rate Prediction}
\label{subsec:atmo_neu_events}

The atmospheric-neutrino interaction rate, which sets the normalization prior to the detector response and single-signal selection, is calculated for flavor~$\alpha$ as:
\begin{equation}
    \begin{aligned}
        N_\alpha = 2\pi TN_A \sum_{t=(\mathrm{C},\mathrm{H})}\frac{m_t}{M_t} \int_{E_{\min}}^{E_{\max}} dE_\nu\int_{-1}^{1} d\cos{\theta_z} \sum_{l=(\nu_e,\nu_\mu)}\phi_l(E_\nu,\cos{\theta_z})P_{l\rightarrow\alpha}(E_\nu,\cos{\theta_z}) \sigma_{\alpha,t}(E_\nu),
    \end{aligned}
\end{equation}
where the factor $2\pi$ results from integration over the azimuthal angle under the assumption of azimuthal symmetry. Here, $\alpha=\nu_e$, $\nu_\mu$, or $\nu_\tau$ denotes the final neutrino flavor, and $T$ is the exposure time, $N_A$ is the Avogadro constant, $m_t$ is the target mass, $M_t$ is the target's molar mass, $\phi_l$ is the differential atmospheric-neutrino flux before oscillation, in units of \si{\per\square\metre\per\second\per\steradian\per\GeV}, $P_{l\rightarrow\alpha}$ is the oscillation probability, and $\sigma_{\alpha,t}(E_\nu)$ is the neutrino-nucleus interaction cross section. For compactness, neutrino and antineutrino modes are implied in the notation and are evaluated separately with their corresponding fluxes, oscillation probabilities, and cross sections. The final single-signal spectra are obtained by applying the detector response and the selection criteria described in Sec.~\ref{subsec:selection}. In this study, we assume a \SI{20}{\kton} LS target composed of 88\% ${}^{12}\mathrm{C}$ and 12\% ${}^{1}\mathrm{H}$ by mass. The integration limits are set to $E_{\min}=\SI{0.1}{\GeV}$ and $E_{\max}=\SI{20}{\GeV}$ for the total rate.

\subsubsection{Atmospheric Neutrino Flux}
\begin{figure}[htbp]
    \centering
    \includegraphics[width=0.48\linewidth]{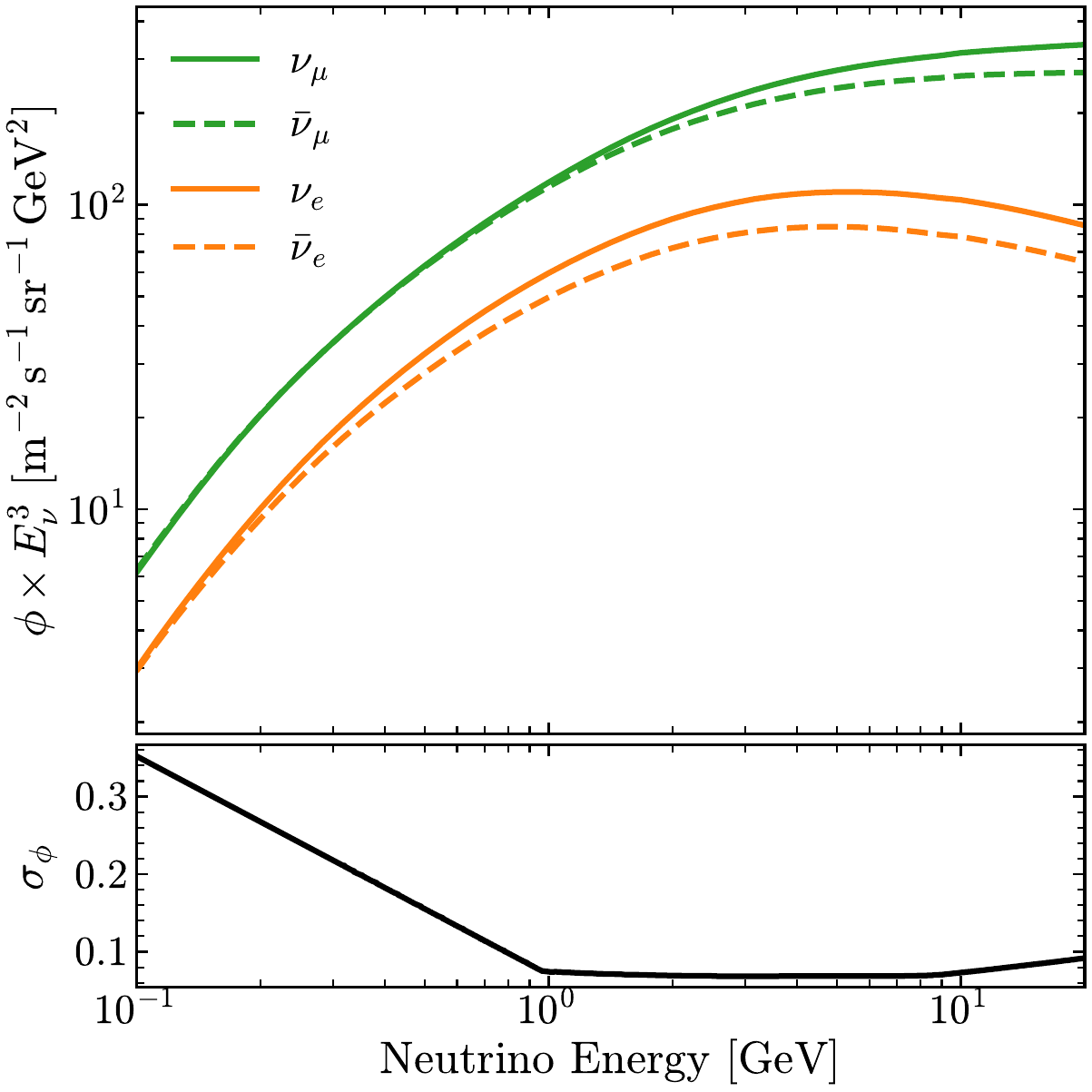}
    \caption{The Honda solar-maximum fluxes for all neutrino flavors at the JUNO site~\cite{honda} (upper panel) and the relative uncertainty~\cite{PhysRevD.75.043006} (lower panel). No neutrino oscillation is considered.}
    \label{fig:flux}
\end{figure}

We use the atmospheric-neutrino flux calculated by the Honda group~\cite{PhysRevD.92.023004}, which is based on a 3D geomagnetic model and the Earth's atmospheric density profile. It provides flux data over a wide energy range, from $\SI{1e-1}{\GeV}$ to $\SI{1e4}{\GeV}$, at various sites including Kamioka, Jiangmen, and the South Pole. Several experiments have confirmed good agreement with these predictions~\cite{Super-Kamiokande:2015qek}. In this study, we use the flux at the JUNO site under solar-maximum conditions. The flux uncertainties are constrained by accurately measured atmospheric muon fluxes~\cite{PhysRevD.100.123022,PhysRevD.75.043006}. Figure~\ref{fig:flux} shows the all-direction-averaged fluxes at the JUNO site together with their associated uncertainties. The relative uncertainty is larger at low energies because of geomagnetic effects and uncertainties in the hadronic interaction model. We note that a recent calculation may reduce the hadronic component of this uncertainty further by exploiting additional atmospheric-muon constraints~\cite{Cheng:2026cqx}.

\subsubsection{Oscillation Probability Calculation}
For NC interactions, oscillations among active flavors do not change the total active neutrino rate, while CC predictions require flavor-dependent oscillation probabilities. Assuming normal hierarchy (NH), we calculate the oscillation probabilities using Prob3++~\cite{prob3plusplus} with the oscillation parameters taken from the Particle Data Group (PDG) 2024 review~\cite{PDG:2024cfk} and the Earth density profile described by the PREM model~\cite{Dziewonski:1981xy}.

\subsubsection{Neutrino Interaction Models}
We consider neutrino energies in the range from $\SI{0.1}{\GeV}$ to $\SI{20}{\GeV}$, since lower-energy events are not well handled by common generators and higher-energy flux contributions are negligible. In the GeV regime, neutrino-nucleus interactions are complex. Different mechanisms dominate in different energy regions and can be grouped into several main categories: quasi-elastic scattering (QE), two-particle-two-hole (2p2h) excitations, resonance production (RES), and deep inelastic scattering (DIS). Coherent scattering is also possible but negligible. We choose the comprehensive model configuration \texttt{G18\_10b\_02\_11b} from GENIE (v3.04.02)~\cite{Andreopoulos:2009rq} as the nominal model because GENIE is widely used by many accelerator-neutrino and atmospheric-neutrino analyses~\cite{MicroBooNE:2019nio,MINERvA:2021csy,IceCubeCollaborationP:2025rpl}. \texttt{G18\_10b\_02\_11b} uses the local Fermi gas model (LFG) to describe the nuclear ground state, with a Fermi momentum that depends on the local nuclear density, in contrast to the relativistic Fermi gas model (RFG)~\cite{GENIE:2021npt}. The model parameters have been tuned by the GENIE collaboration~\cite{GENIE:2021zuu}. In this configuration, the quasi-elastic interaction is described by the Valencia model~\cite{Nieves:2004wx} and the two-particle-two-hole contribution by the model of Nieves, Ruiz Simo, and Vicente Vacas~\cite{Nieves:2011pp}, while resonance production and deep inelastic scattering are described by the Berger--Sehgal~\cite{PhysRevD.76.113004} and Bodek--Yang~\cite{BODEK200270} models, respectively, and the final-state interactions by the hN intranuclear cascade~\cite{Serber:1947zza,Metropolis:1958wvo,Metropolis:1958sb}. To assess the model dependence, alternative configurations with different nuclear models, FSI treatments, generators, and de-excitation and secondary-interaction models are also considered, as summarized in Table~\ref{tab:models} and discussed in Secs.~\ref{sec:processes} and \ref{sec:result}.

\begin{figure}[htbp]
    \centering
    \includegraphics[width=0.48\linewidth]{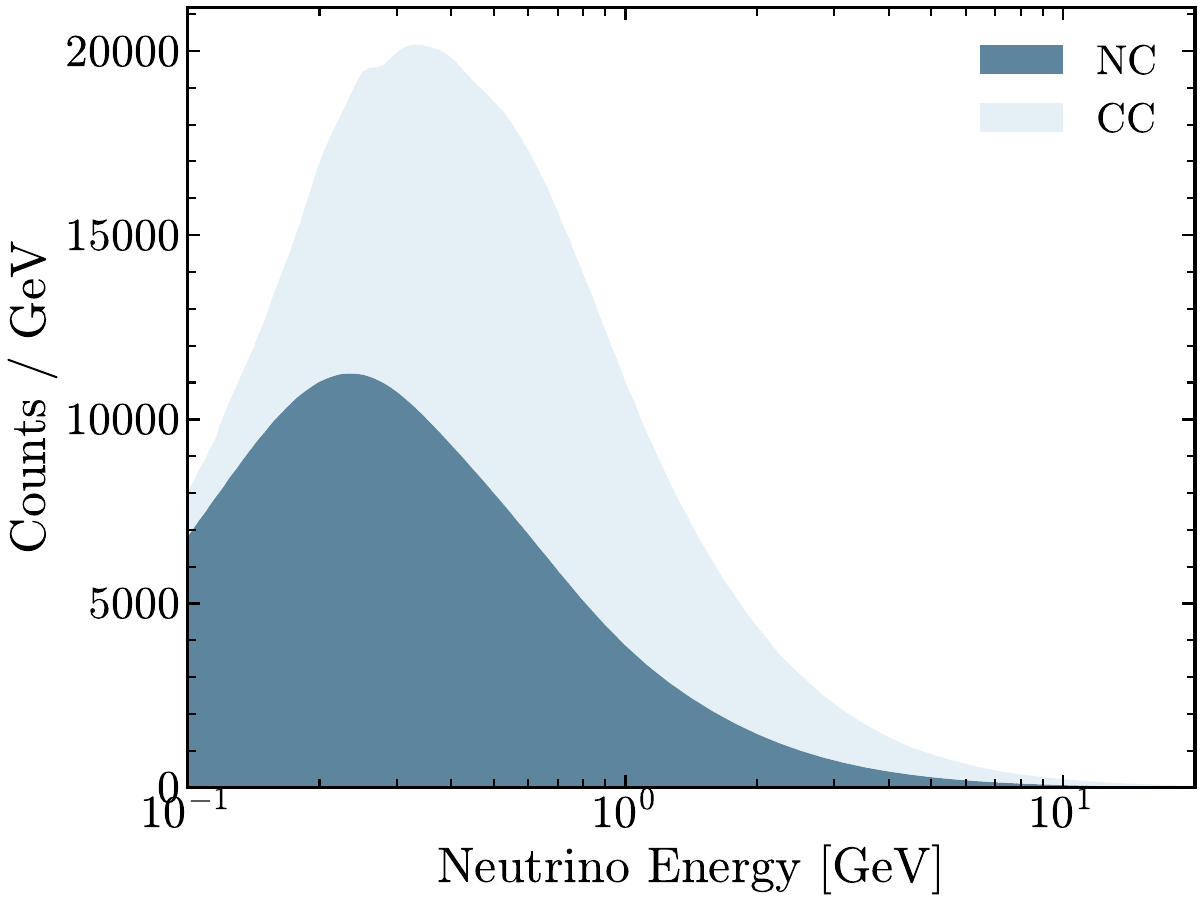}
    \caption{Incident-neutrino energy distributions of atmospheric-neutrino events in a \SI{20}{\kton} LS detector for a \SI{10}{\yr} exposure based on the GENIE \texttt{G18\_10b\_02\_11b} model.}
    \label{fig:rate}
\end{figure}

The incident-neutrino energy distributions for the nominal model, separated into NC and CC components, are presented in Figure~\ref{fig:rate}. Since the atmospheric-neutrino fluxes approximately follow a power-law distribution, atmospheric-neutrino interactions are predominantly in the sub-GeV region, especially for NC events. This energy range is therefore the main focus for the systematic comparison of neutrino-nucleus interaction models discussed in Sec.~\ref{sec:processes}.

\subsubsection{Residual-Nucleus De-Excitation}
After FSI, the residual nucleus may contain one or more nucleon holes and can be left in an excited state. During de-excitation, additional particles such as $\gamma$ rays, neutrons, protons, or $\alpha$ particles may be emitted. These emitted particles affect the single-signal selection in two distinct ways. First, they modify the delayed-signal topology: the emitted neutrons directly lead to neutron-capture signals, while the $\gamma$ rays, protons, and $\alpha$ particles can generate additional neutrons through secondary interactions, introducing additional neutron-capture tags and thereby migrating otherwise single-signal events into neutron-tagged categories. Second, they affect the prompt energy spectrum: the emitted particles deposit energy in the LS, with the emitted protons contributing to the prompt visible energy, thereby modifying the final spectrum observed by the LS detector.

\begin{table}
    \centering
    \caption{The shell-hole configurations, excitation energies, and configuration probabilities for the excited residual nuclei produced in ${}^{12}\mathrm{C}$ interactions. Ground-state configurations are not listed.}
    \begin{tabular}{c c c c}
        \toprule
        Residual Nucleus & Shell Hole & $E_x$ & Configuration Probability\\
        \midrule
         ${}^{11}\mathrm{C}^{*}/{}^{11}\mathrm{B}^{*}$ & $s_{1/2}$ & $\SI{23}{\MeV}$ & $1/3$ \\
         \multirow{2}{*}{${}^{10}\mathrm{C}^{*}/{}^{10}\mathrm{Be}^{*}$} & $s_{1/2}$ & $\SI{46}{\MeV}$ & $1/15$ \\
            & $s_{1/2} \& p_{3/2}$ & $\SI{23}{\MeV}$ & $8/15$ \\
        \multirow{2}{*}{${}^{10}\mathrm{B}^{*}$} & $s_{1/2}$ & $\SI{46}{\MeV}$ & $1/9$ \\
            & $s_{1/2} \& p_{3/2}$ & $\SI{23}{\MeV}$ & $4/9$ \\
         \bottomrule
    \end{tabular}
    \label{tab:deex}
\end{table}

Recent studies of the atmospheric NC background for DSNB searches have shown that de-excitation cannot be ignored in experiments, e.g., Super-K~\cite{Super-Kamiokande:2021jaq}, KamLAND~\cite{KamLAND:2022ptk}, and JUNO~\cite{Cheng:2020aaw,Cheng:2020oko}. However, this process is not implemented in standard neutrino event generators, and the residual-nucleus excitation energy needed as input to an external de-excitation model is usually not available. To assign the residual-nucleus excitation energy, we choose the simple shell model of the ${}^{12}\mathrm{C}$ ground state~\cite{Kamyshkov:2002wp}, with two nucleons in the $s_{1/2}$ shell, four nucleons in the $p_{3/2}$ shell, and no nucleon in the $p_{1/2}$ shell. If nucleons are knocked out from the $p_{3/2}$ shell, the residual nucleus is assumed to be in the ground state. The knock-out of a nucleon from the $s_{1/2}$ shell results in an excited state of the residual nucleus. All residual nuclei with $Z\geq3$ and $A\geq7$ are considered, and Table~\ref{tab:deex} shows the cases where one or two nucleons are knocked out from ${}^{12}\mathrm{C}$. With the nuclide and excitation energy specified, TALYS~\cite{Koning:2005ezu} is used as the nominal de-excitation model to simulate the de-excitation channels and provide the emitted particles and their kinetic energies.

\subsubsection{Secondary Interactions in LS}
\begin{figure}[htbp]
    \centering
    \includegraphics[width=0.48\linewidth]{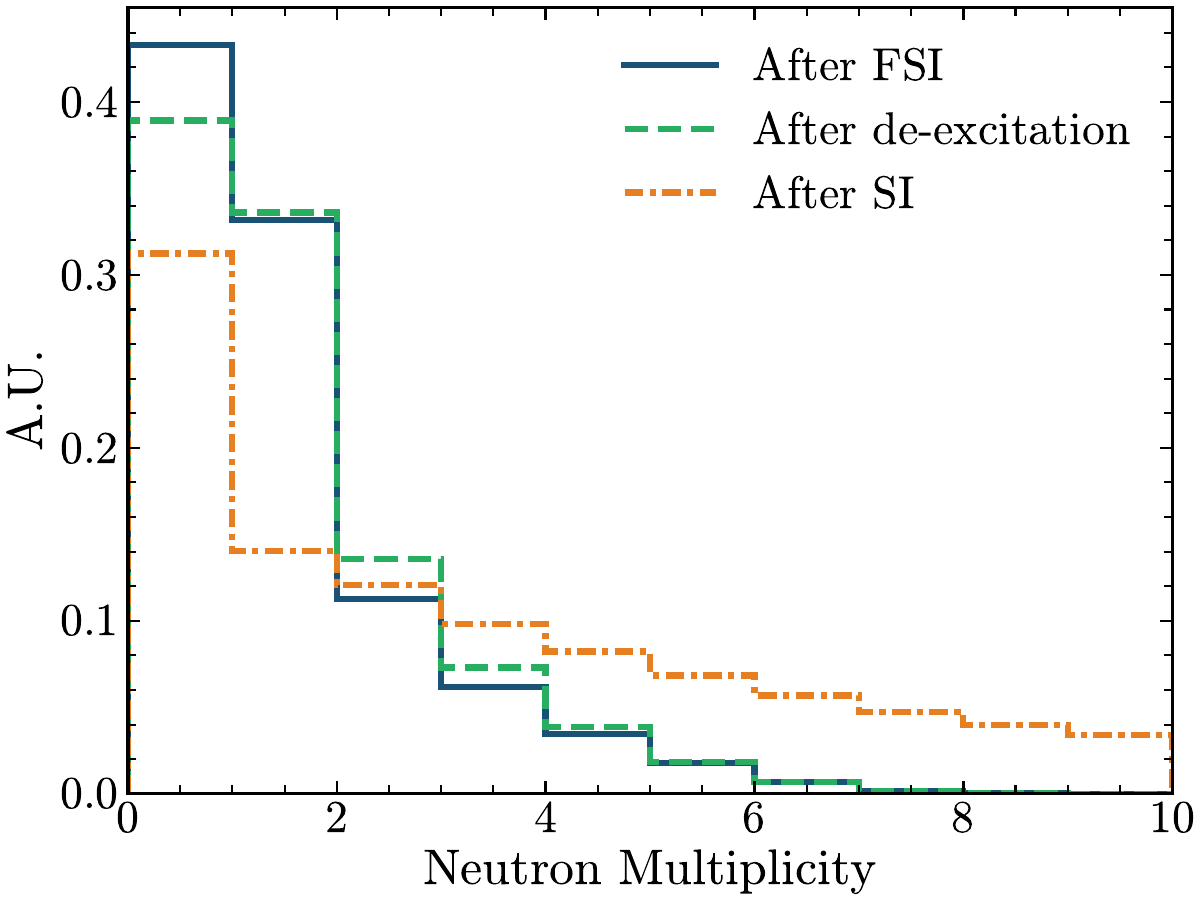}
    \includegraphics[width=0.48\linewidth]{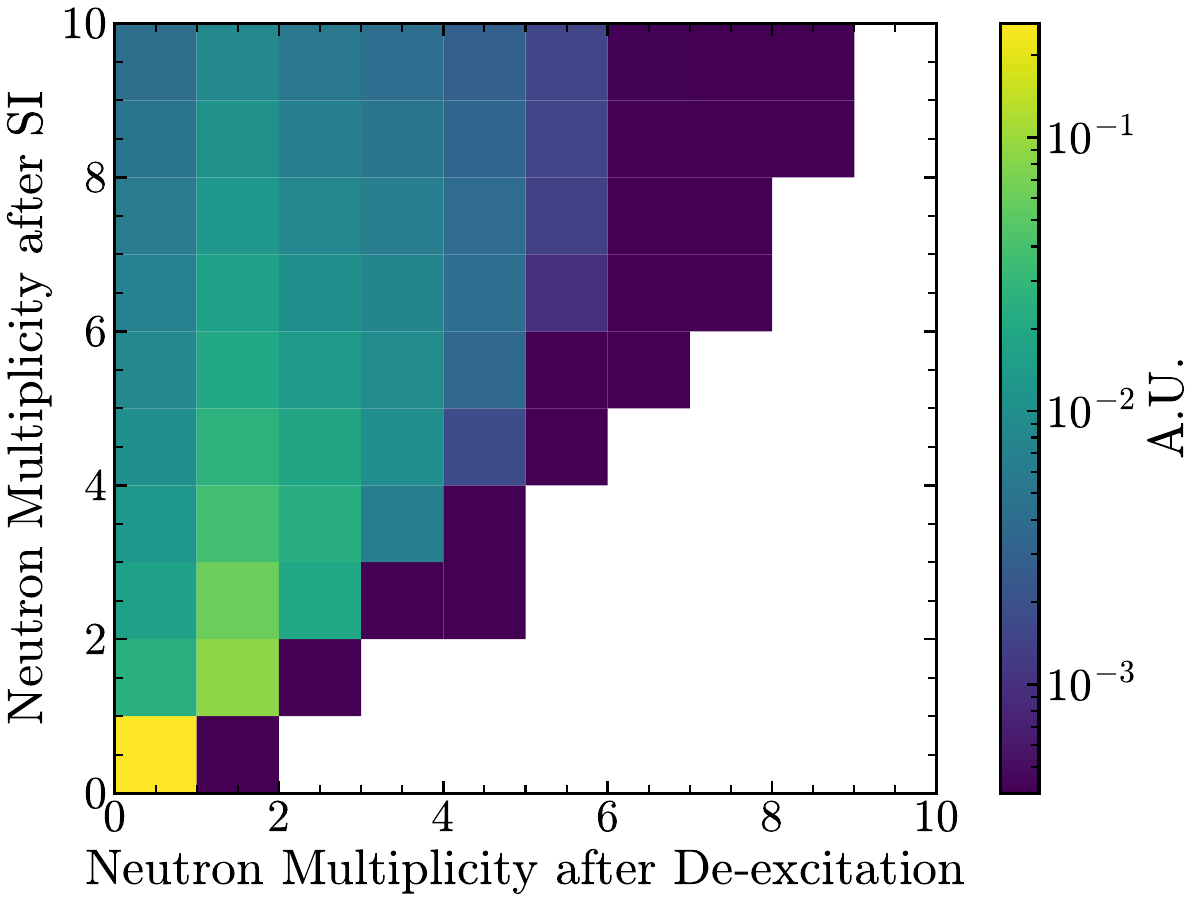}
    \caption{Evolution of neutron multiplicity in atmospheric-neutrino interactions in LS, with both panels normalized to unit area. (left) 1D neutron multiplicity distributions at different interaction stages; (right) 2D neutron multiplicity distributions before and after SI.}
    \label{fig:si}
\end{figure}

The final step in predicting detectable events is to consider particle propagation in LS detectors. After the primary neutrino-nucleus interaction, FSI, and residual-nucleus de-excitation, particles may undergo secondary interactions with ${}^{12}\mathrm{C}$ or ${}^{1}\mathrm{H}$ in the LS. Therefore, both the visible-energy spectrum and the number of delayed signals, such as neutron-capture signals and decay products, strongly depend on SI models. The evolution of the neutron multiplicity through the interaction chain is illustrated in Figure~\ref{fig:si}. The multiplicity after de-excitation differs from that after FSI because excited residual nuclei can emit additional neutrons, while the final captured-neutron multiplicity can differ further after particle transport in LS. During SI, outgoing particles, such as neutrons, can generate additional neutrons through inelastic scattering, leading to an increased neutron multiplicity. Conversely, some neutrons may be absorbed during inelastic scattering, resulting in a decreased neutron multiplicity.

We use GEANT4~\cite{GEANT4:2002zbu} version 11.4 to simulate these processes in a detector consisting of an LS sphere of radius $\SI{17.7}{\metre}$ at the center of a water cylinder. The LS is composed of 88\% ${}^{12}\mathrm{C}$ and 12\% ${}^{1}\mathrm{H}$ by mass. GEANT4's \texttt{FTFP\_BERT\_HP} physics list, which combines the Fritiof string model, the Bertini cascade, and high-precision neutron transport, is used as the nominal model when discussing the selection in Sec.~\ref{subsec:selection} and predicting the single-signal event rate and energy spectra in Sec.~\ref{subsec:rate_spectra}.

\subsection{Selection of Singles}
\label{subsec:selection}

The visible energy $E_{\mathrm{vis}}$ is defined as the energy deposited in the LS after applying the quenching response described by Birks' law, using the parameters from Ref.~\cite{JUNO:2015zny}. The prompt energy $E_{\mathrm{prompt}}$ is defined as the total visible energy deposited in the first \SI{1008}{\ns} interval, corresponding to $t\in[0,\SI{1008}{\ns})$. This time window contains essentially the full prompt scintillation light of a signal.

In this work, we adopt an idealized detector response that includes GEANT4-based particle transport together with the quenching and timing response of the LS, but neglects position-dependent energy reconstruction, energy-resolution smearing, trigger efficiency, and delayed-signal tagging efficiencies. The signal definition and event selection therefore operate directly on the simulated energy deposits, and all predictions presented in this paper should be regarded as ideal-detector expectations rather than a quantitative description of the data.

To select singles, we require no delayed signal after the prompt window and apply the following cuts sequentially:
\begin{enumerate}
    \item Prompt-energy threshold: $E_{\mathrm{prompt}}\geq\SI{0.7}{\MeV}$, set following the JUNO reactor-oscillation analysis~\cite{JUNO:2025gmd}.
    \item Fully contained (FC) cut: no visible energy leakage from the LS volume into the surrounding water. Experimentally, this corresponds to requiring no significant Cherenkov activity in the water-Cherenkov veto.
    \item Neutron-capture veto: no signal with $E_{\mathrm{vis}}$ in either the $[1.9,2.5]$~\si{\MeV} or $[4.4,5.5]$~\si{\MeV} energy window, corresponding to neutron capture on hydrogen and carbon, respectively, in the time interval $t\in[\SI{1008}{\ns},\SI{1000}{\us}]$.
    \item Michel-electron veto: no signal with $E_{\mathrm{vis}}\in[6,53]$~\si{\MeV} in the time interval $t\in[\SI{1008}{\ns},\SI{10}{\us}]$.
\end{enumerate}

\begin{figure}[htbp]
    \centering
    \includegraphics[width=0.95\linewidth]{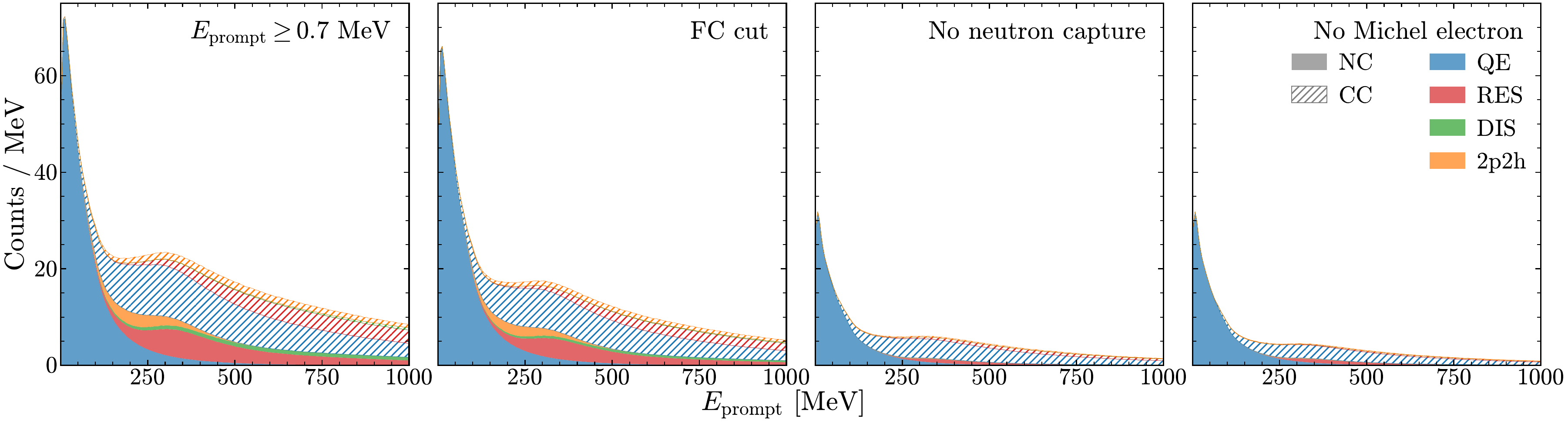}
    \caption{Prompt-energy spectra after the sequential selection cuts for a \SI{20}{\kton} LS detector with a \SI{10}{\yr} exposure. NC and CC components are drawn in the same panels: NC contributions are shown as filled stacked histograms, while CC contributions are overlaid with hatching. The colors separate the interaction processes QE, RES, DIS, and 2p2h.}
    \label{fig:sel}
\end{figure}

Delayed signals are searched for only after the first \SI{1008}{\ns} interval. This is a conservative choice, since prompt and delayed components occurring within the same time window could, in principle, be separated using their different time profiles. Based on the event rates calculated in Sec.~\ref{subsec:atmo_neu_events}, the $E_{\mathrm{prompt}}$ spectra after each selection step are shown in Figure~\ref{fig:sel}. The threshold and FC cuts define the prompt fully contained sample, while the neutron-capture and Michel-electron vetoes suppress events with correlated delayed signals.

The selected sample is expected to be dominated by the interaction topologies listed below, which contain no delayed signal within the veto windows. Due to the finite time and energy acceptance of the delayed-signal vetoes, interactions containing neutrons or charged pions can also pass the selection when their delayed signals are not identified.
\begin{itemize}
    \item NCQE: $\overset{(-)}{\nu} + {}^{1}\mathrm{H} \rightarrow \overset{(-)}{\nu} + \mathrm{p}$ and $\overset{(-)}{\nu} + {}^{12}\mathrm{C} \rightarrow \overset{(-)}{\nu} + N\mathrm{p} + \mathrm{X^{(*)}}$
    \item CCQE: $\overset{(-)}{\nu} + {}^{12}\mathrm{C} \rightarrow l^{\mp} + N\mathrm{p} + \mathrm{X^{(*)}}$
\end{itemize}
where $N$ is the number of outgoing protons ($N>0$), and $\mathrm{X^{(*)}}$ represents the residual nucleus, which may be in an excited state. The 2p2h processes can also produce similar topologies, although the contribution is subdominant in the selected sample. Since $\pi^{0}$ decays immediately, RES and DIS also contribute to singles:
\begin{itemize}
    \item NCRES/DIS: $\overset{(-)}{\nu} + {}^{1}\mathrm{H} \rightarrow \overset{(-)}{\nu} + \mathrm{p} + \pi^{0}$ and $\overset{(-)}{\nu} + {}^{12}\mathrm{C} \rightarrow \overset{(-)}{\nu} + N\mathrm{p} + \pi^{0} + \mathrm{X^{(*)}}$
    \item CCRES/DIS: $\overset{(-)}{\nu} + {}^{12}\mathrm{C} \rightarrow l^{\mp} + N\mathrm{p} + \pi^{0} + \mathrm{X^{(*)}}$
\end{itemize}

Under the nominal model and the selection criteria described above, approximately 12498 NC and 19347 CC atmospheric-neutrino interactions for an exposure of \SI{200}{\kton\cdot\yr} are reduced to approximately 2907 NC and 2296 CC singles over the full prompt-energy range, including events with $E_{\mathrm{prompt}}\geq\SI{1000}{\MeV}$. NCQE and CCQE are the dominant interaction channels for the NC and CC samples, respectively. The relative NC/CC composition varies strongly with the prompt energy: NC events dominate the sample at low energies, especially below about \SI{100}{\MeV}, while CC events become dominant at higher energies.

\section{Systematic Effects on Single-Signal Events}
\label{sec:processes}

\subsection{Neutrino-Nucleus Interaction}

The first source of model dependence is the primary atmospheric-neutrino interaction. In addition to GENIE, we use the NuWro Monte Carlo generator (version 21.09.2)~\cite{Juszczak:2005zs} to compare the impact of different theoretical descriptions of neutrino-nucleus scattering. In LS, the primary targets are ${}^{12}\mathrm{C}$ and ${}^{1}\mathrm{H}$. The hydrogen target is a free proton, and its NCQE scattering is described by the Ahrens model~\cite{Ahrens:1986xe}, which gives the differential cross section:
\begin{equation}
\label{ahren}
    \frac{d\sigma}{dQ^2}=\frac{G_F^2 M_p^2}{8\pi E_\nu^2}\left(A\pm B\frac{(s-u)}{M_p^2}+C\frac{(s-u)^2}{M_p^4}\right),
\end{equation}
where the plus (minus) sign applies to neutrinos (antineutrinos), $s-u=4M_p E_\nu-Q^2$, and $A$, $B$, and $C$ are form-factor combinations that depend on the momentum transfer. For elastic scattering on a free proton, $Q^2=2M_pT_p$, where $T_p$ is the proton kinetic energy. More details can be found in Ref.~\cite{Ahrens:1986xe}. For NCQE scattering on ${}^{1}\mathrm{H}$, the differences between GENIE and NuWro arise mainly from parameter choices such as the axial mass $M_A$ (Table~\ref{tab:models}); the comparison between the two generators therefore already incorporates part of the elementary neutrino-nucleon form-factor variation. As a result, the generator dependence for NCQE interactions on hydrogen is relatively small within the tested configurations.

For ${}^{12}\mathrm{C}$, the nucleons are bound and nuclear effects substantially modify the free-nucleon picture. The impulse approximation (IA)~\cite{Frullani:1984nn} is commonly used to describe the nucleus in terms of individual quasi-free nucleons. GENIE implements both the RFG and LFG descriptions, with the former including the Bodek-Ritchie (BRRFG) variant~\cite{Bodek:1981wr}, while NuWro additionally provides the spectral function (SF) model~\cite{Benhar:1989aw}. These nuclear-model choices affect the outgoing nucleon multiplicity and energy distribution, and hence both the prompt-energy spectrum and the probability of passing the delayed-signal vetoes.

\begin{figure}[htbp]
    \centering
    \begin{subfigure}{0.48\linewidth}
        \centering
        \includegraphics[width=\linewidth]{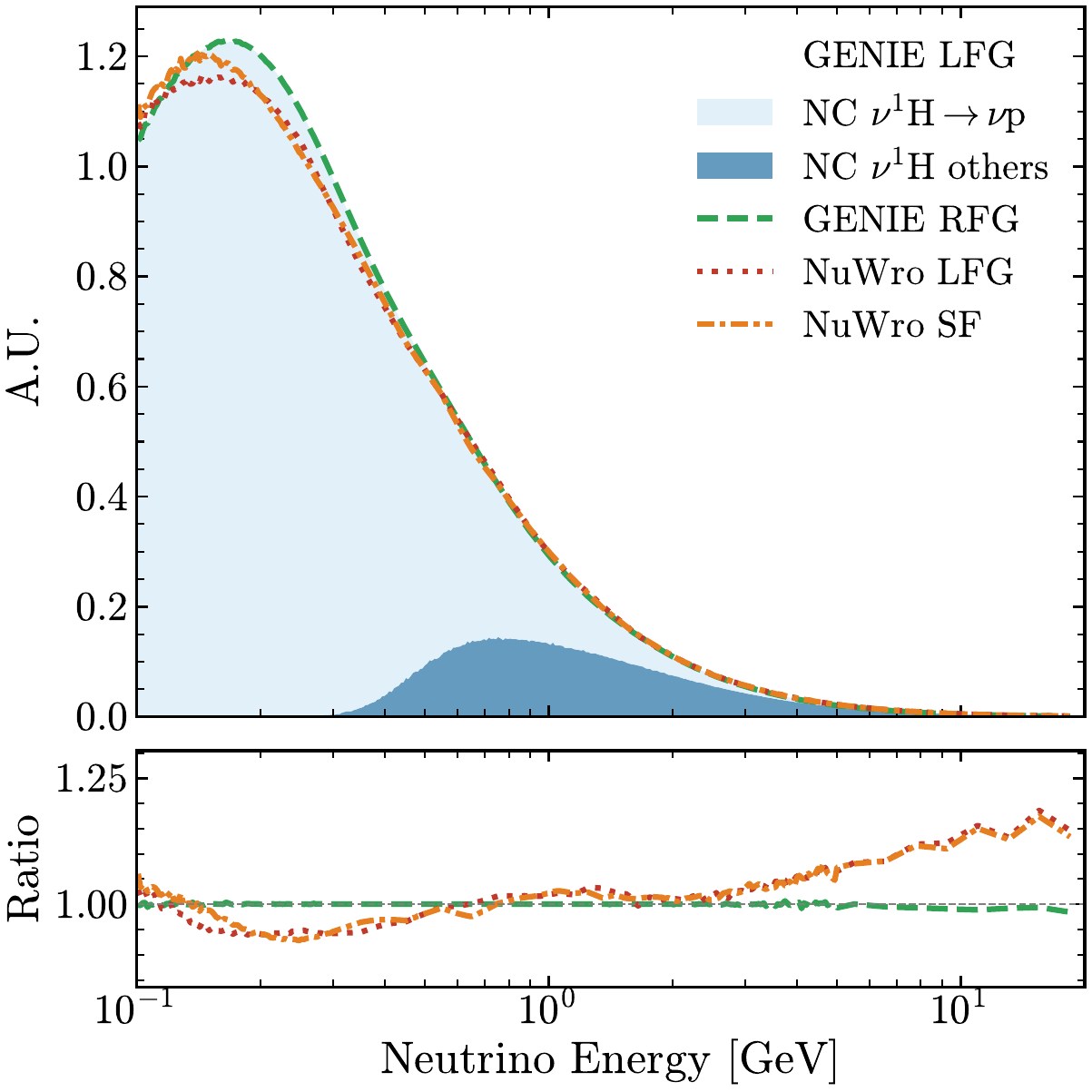}
        \caption{NC interactions on ${}^{1}\mathrm{H}$.}
    \end{subfigure}
    \hfill
    \begin{subfigure}{0.48\linewidth}
        \centering
        \includegraphics[width=\linewidth]{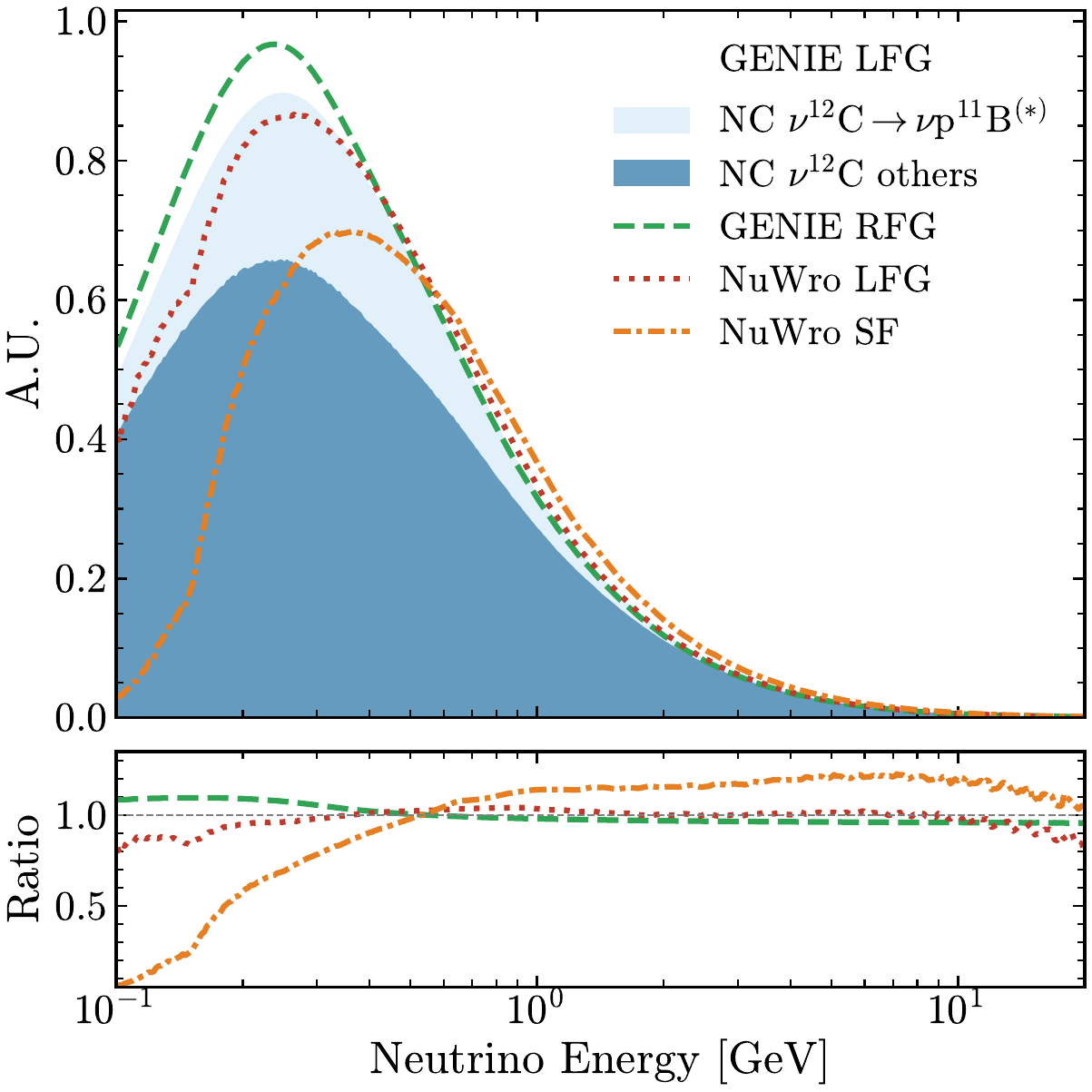}
        \caption{NC interactions on ${}^{12}\mathrm{C}$.}
    \end{subfigure}
    \caption{Incident-neutrino energy distributions for NC interactions on hydrogen and carbon targets. The upper-panel distributions in both subfigures are normalized by their integrals and bin widths. The stacked histograms show the nominal GENIE LFG prediction, decomposed into the leading channel and the remaining NC contribution for each target. The overlaid curves compare alternative neutrino-interaction models, and the lower panels show their ratios to the nominal prediction.}
    \label{fig:enu_nc_components}
\end{figure}

The impact of these model choices on NC interactions is examined separately for hydrogen and carbon targets in Figure~\ref{fig:enu_nc_components}. In the nominal model, the dominant hydrogen contribution is the NCQE channel $\nu+{}^{1}\mathrm{H}\rightarrow\nu+\mathrm{p}$, while the carbon spectrum contains the exclusive single-proton channel $\nu+{}^{12}\mathrm{C}\rightarrow\nu+\mathrm{p}+{}^{11}\mathrm{B}^{(*)}$ together with other NC carbon final states. The NC incident-energy spectrum varies in shape across generator and nuclear-model choices. The model spread is much larger for carbon than for hydrogen, especially in the sub-GeV region, because carbon predictions depend on the nuclear ground-state model, nuclear correlations, and FSI treatment in addition to the elementary neutrino-nucleon interaction, within the set of models considered here. The NuWro SF prediction differs most strongly from the other models, predicting fewer low-energy NC events but relatively more events at higher energies.

After the primary interaction, the produced hadrons may undergo FSI inside the nucleus, including elastic and inelastic scattering, pion production, hadron absorption, and charge exchange. These processes can change the neutron and proton multiplicities before the particles enter the LS, making FSI particularly relevant for single-signal selection. We compare three GENIE FSI treatments: hA, hN, and G4B. The hA model is empirical, whereas hN is a full intranuclear cascade (INC) model~\cite{Serber:1947zza,Metropolis:1958wvo,Metropolis:1958sb}, similar in spirit to the INC treatment in NuWro. The default NuWro cascade treatment is referred to as Csc in the model comparisons below. The G4B model is the GEANT4 Bertini cascade used in the alternative GENIE \texttt{G18\_10d\_02\_11b} configuration. For CCQE interactions, we compare the Valencia model~\cite{Nieves:2004wx}, which is used in the nominal GENIE \texttt{G18\_10b\_02\_11b} configuration and includes LFG, Coulomb correction, and Random Phase Approximation (RPA) effects, with the Llewellyn-Smith model~\cite{LlewellynSmith:1971uhs} used in the alternative configurations listed in Table~\ref{tab:models}.

\begin{figure}[htbp]
    \centering
    \begin{subfigure}{0.48\linewidth}
        \centering
        \includegraphics[width=\linewidth]{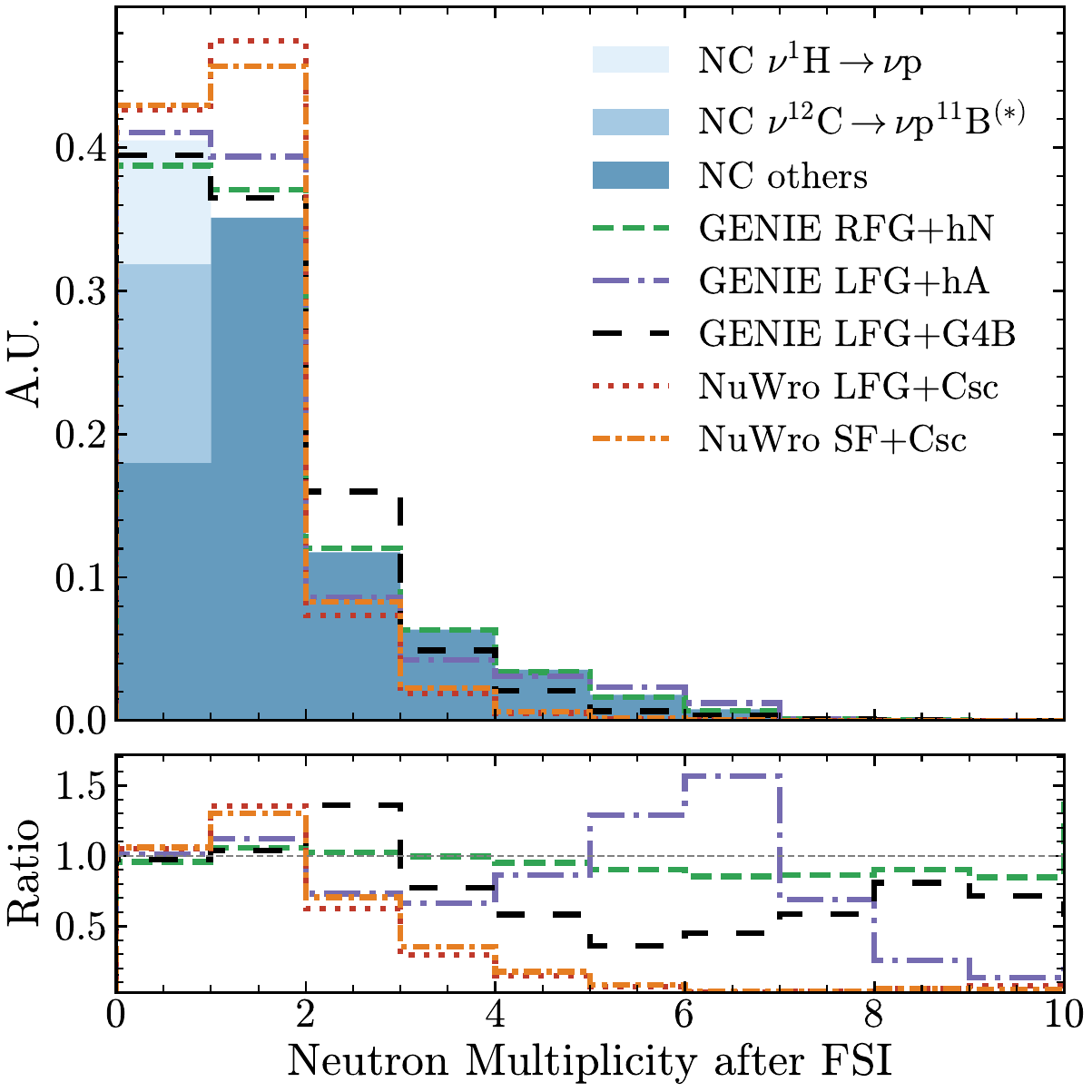}
    \end{subfigure}
    \hfill
    \begin{subfigure}{0.48\linewidth}
        \centering
        \includegraphics[width=\linewidth]{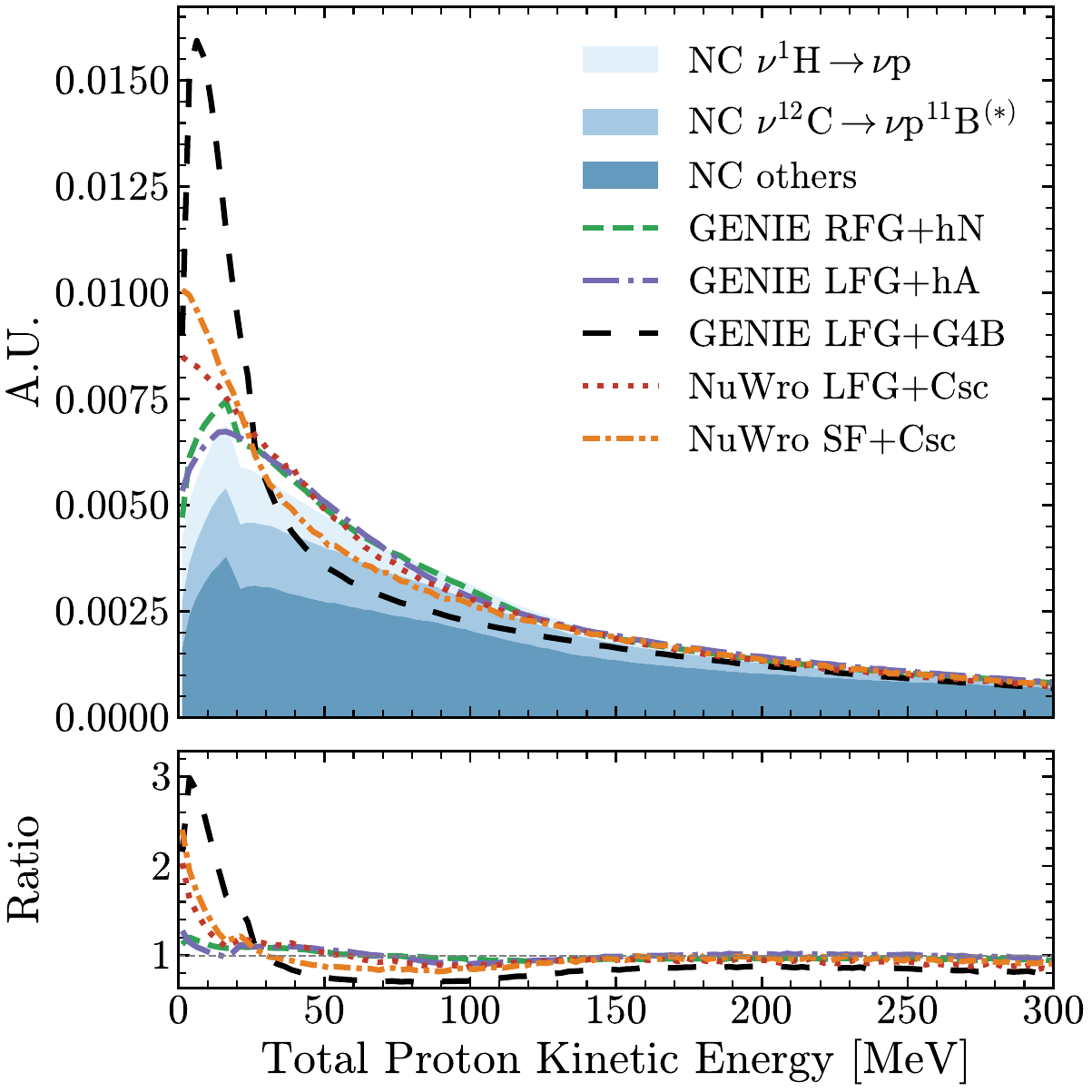}
    \end{subfigure}
    \caption{Comparison of final-state nucleon observables after FSI for NC interactions in atmospheric-neutrino simulations: (left) neutron multiplicity; (right) event-by-event sum of the kinetic energies of all final-state protons. The upper-panel distributions in both subfigures are normalized to unit area. The nominal GENIE LFG prediction is stacked into hydrogen, the leading carbon single-proton channel, and the remaining NC contribution, and the overlaid curves compare alternative nuclear-model and FSI choices.}
    \label{fig:nmult_nc}
\end{figure}

The effects of these nuclear-model and FSI choices on the final-state nucleon content before de-excitation are summarized in Figure~\ref{fig:nmult_nc}. For the neutron multiplicity, the alternative curves show that the distribution changes noticeably across models, but the $N=0$ category is comparatively stable: the largest relative differences appear in the $N\geq1$ region, where neutron emission is more sensitive to the nuclear ground-state model and to FSI. The same model choices also affect the total kinetic energy carried by final-state protons. The differences are most pronounced below about \SI{30}{\MeV}; in particular, the G4B FSI model produces a much stronger peak around \SI{10}{\MeV}, which is consistent with greater fragmentation of the residual nucleus and less kinetic energy remaining in the outgoing protons. This behavior is relevant for the single-signal sample, because the selected events are dominated by neutron-free topologies, while the visible prompt energy is largely determined by the proton kinetic-energy spectrum after quenching.

\subsection{De-Excitation}

\begin{figure}[htbp]
    \centering
    \includegraphics[width=0.48\linewidth]{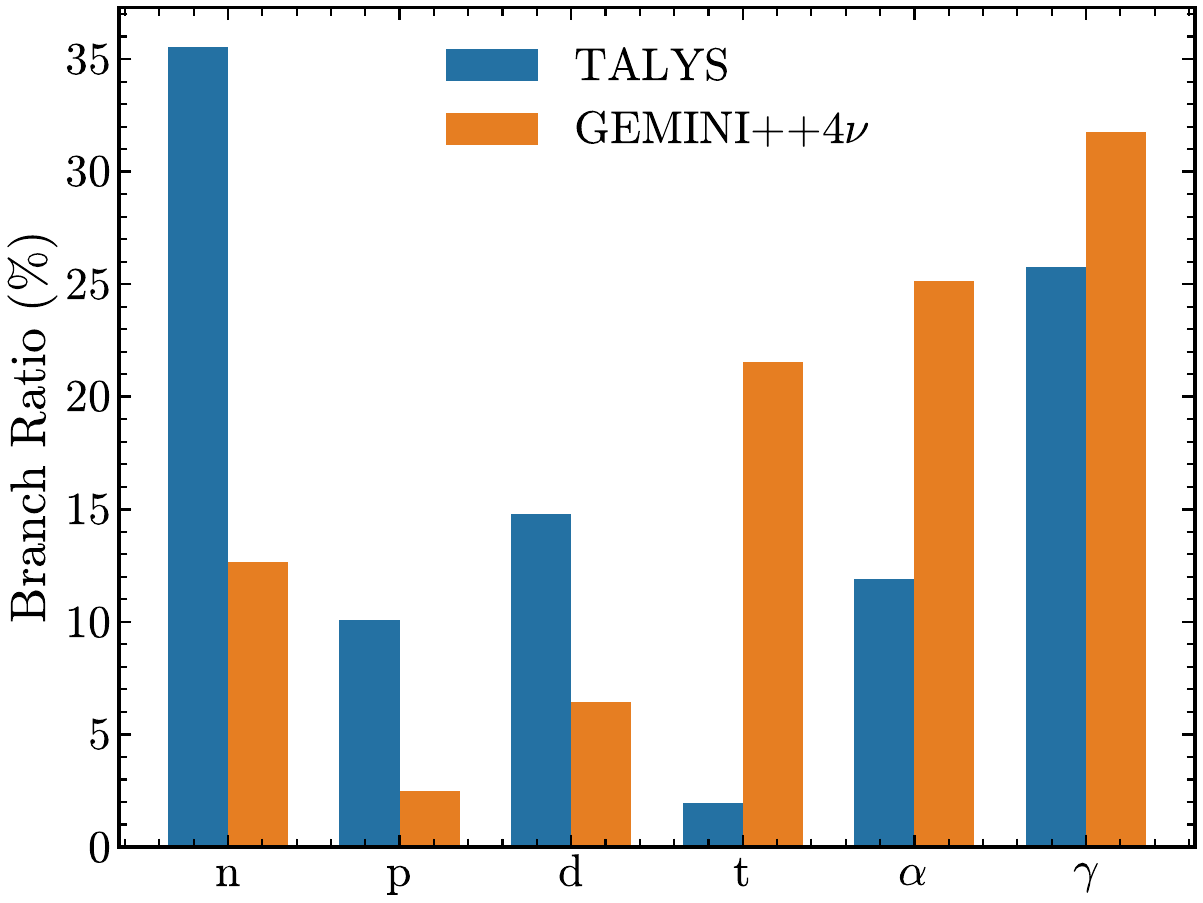}
    \caption{Comparison of the predicted branching ratios for $\mathrm{n}$, $\mathrm{p}$, $\mathrm{d}$, $\mathrm{t}$, $\alpha$, and $\gamma$ emissions from the de-excitation of ${}^{11}\mathrm{B}^{*}$ at $E_x = \SI{23}{\MeV}$. GEMINI++4$\nu$ predicts more heavy charged particles, whereas TALYS predicts more neutron emission.}
    \label{fig:deex}
\end{figure}

We compare two de-excitation treatments, TALYS and GEMINI++4$\nu$~\cite{Niu:2024hgs}. TALYS serves as the nominal model in this work; it is widely used to calculate branching ratios and emitted-particle spectra for nuclear de-excitation channels. GEMINI++4$\nu$ is an alternative treatment modified from GEMINI++~\cite{PhysRevC.82.014610,PhysRevC.82.044610} for the de-excitation of highly excited light residual nuclei, such as the ${}^{11}\mathrm{B}^{*}$ produced in liquid scintillator detectors. For ${}^{11}\mathrm{B}^{*}$, it provides the best agreement with nuclear experimental data among several widely used statistical-model codes. Since ${}^{11}\mathrm{B}^{*}$ is exactly the residual nucleus produced in the single-proton NC channel on ${}^{12}\mathrm{C}$, we adopt GEMINI++4$\nu$ to estimate the model dependence of the de-excitation process. The predicted branching ratios for ${}^{11}\mathrm{B}^{*}$ de-excitation at $E_x=\SI{23}{\MeV}$ are compared in Figure~\ref{fig:deex}. Relative to TALYS, GEMINI++4$\nu$ predicts fewer emitted neutrons and more heavy charged particles, which directly affects the neutron-veto efficiency and the selected single-signal rate.

\begin{figure}[htbp]
    \centering
    \begin{subfigure}{0.48\linewidth}
        \centering
        \includegraphics[width=\linewidth]{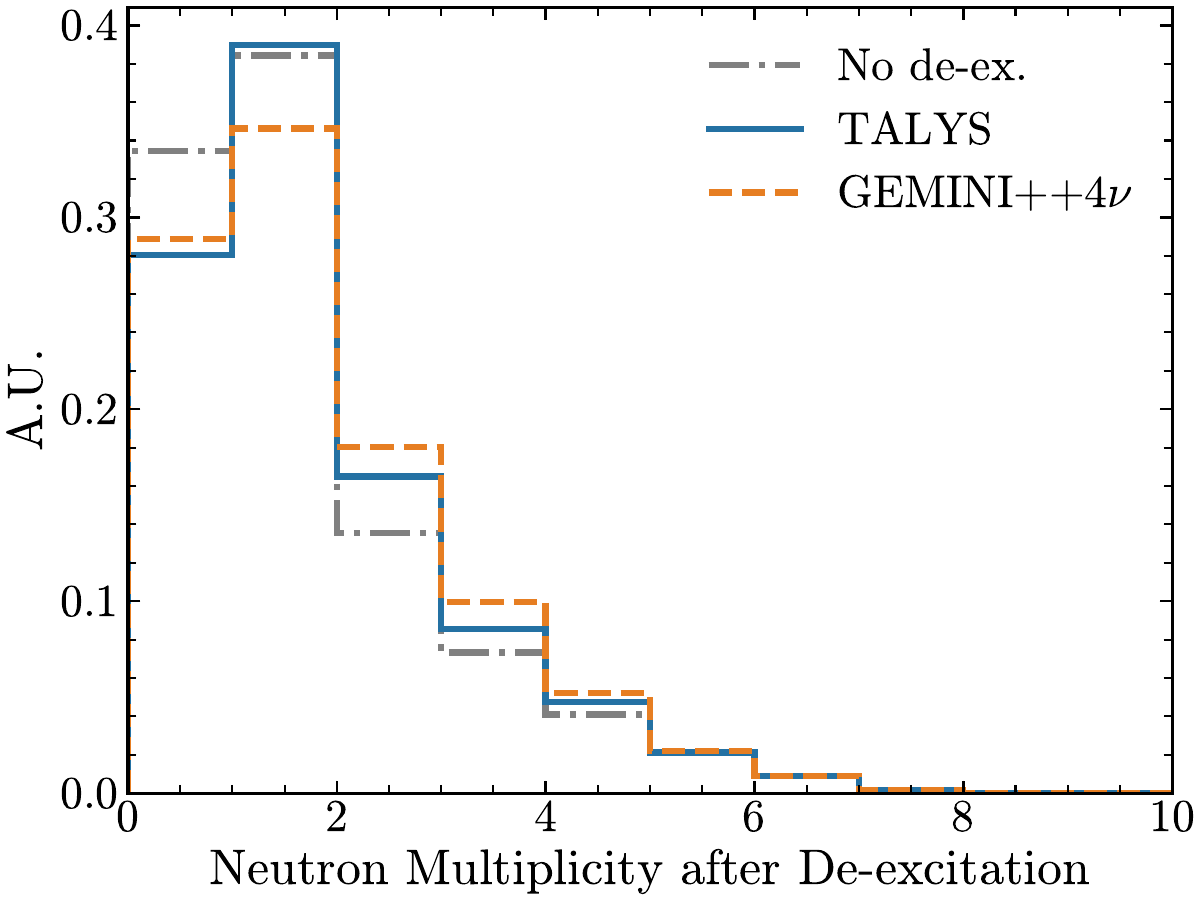}
    \end{subfigure}
    \hfill
    \begin{subfigure}{0.48\linewidth}
        \centering
        \includegraphics[width=\linewidth]{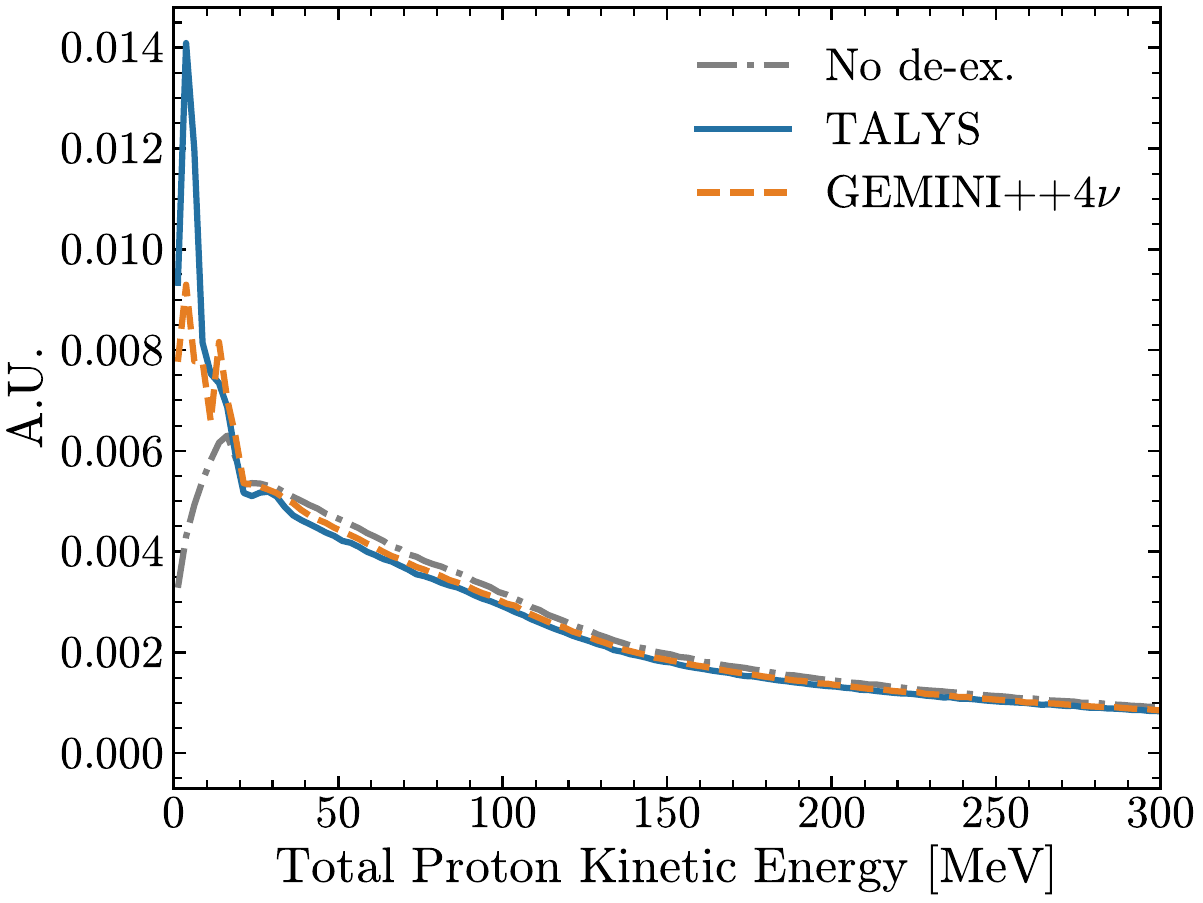}
    \end{subfigure}
    \caption{Effects of de-excitation on atmospheric-neutrino NC interactions with ${}^{12}\mathrm{C}$. The distributions in both subfigures are normalized to unit area. (left) compares the neutron multiplicity after de-excitation, showing that neutron emission from excited residual nuclei reduces the number of events without neutrons; (right) compares the total proton kinetic energy after de-excitation, where the extra protons mainly populate the \SIrange{0}{20}{\MeV} region and therefore affect the low-energy visible spectrum.}
    \label{fig:deex_effects}
\end{figure}

For each event, the residual nucleus type from the generator and the sampled excitation state are used as inputs to the de-excitation calculation. The effect of de-excitation on the neutron multiplicity and the total proton kinetic energy in $\nu{}^{12}\mathrm{C}$ NC interactions is summarized in Figure~\ref{fig:deex_effects}. Events that contain no outgoing neutron before de-excitation can acquire additional neutrons from the excited residual nucleus and are then rejected by the neutron-capture veto. TALYS and GEMINI++4$\nu$ give different estimates of this additional neutron emission, but the discrepancy is smaller for the $N=0$ category than for $N=1$. This is favorable for the single-signal sample, because the selected singles are dominated by neutron-free topologies and therefore inherit a reduced model dependence from de-excitation. In contrast, NC events contributing to the inverse-beta-decay (IBD)-like background of DSNB searches rely more strongly on the $N=1$ component, so the de-excitation uncertainty is more relevant there. Overall, the small model spread in the singles sample helps reduce the corresponding model-induced systematic uncertainty. The same de-excitation stage also emits protons, and these extra protons mainly populate the low-energy part of the visible spectrum. The additional proton kinetic energy is concentrated below about \SI{20}{\MeV}, so the corresponding impact on the final visible-energy spectrum is also expected to be strongest in this low-energy region.

\subsection{Secondary Interaction}

After leaving the primary nucleus, the outgoing particles propagate through the LS and may undergo SI with ${}^{12}\mathrm{C}$ or ${}^{1}\mathrm{H}$. In GEANT4, the hadronic part of a physics list describes the relevant elastic, inelastic, and capture processes. The nominal physics list is \texttt{FTFP\_BERT\_HP}, referred to as BERT hereafter. We also compare \texttt{QGSP\_BIC\_HP} and \texttt{QGSP\_INCLXX\_HP}, referred to as BIC and INCL++, respectively. In addition, we include \texttt{FTFP\_BERT\_HP} with the precompound model enabled, denoted as \texttt{FTFP\_BERT\_HP+PC}, or BERT+PC hereafter. Here, PC denotes the GEANT4 Precompound model used for nuclear de-excitation after the cascade stage~\cite{QuesadaMolina:2011pc}. The HP suffix indicates that neutrons with energies of \SI{20}{\MeV} and below are transported with the high-precision neutron models and cross-section data for elastic scattering, inelastic scattering, capture, and fission:
\begin{itemize}
    \item BERT: \texttt{FTFP\_BERT\_HP}
    \item BIC: \texttt{QGSP\_BIC\_HP}
    \item INCL++: \texttt{QGSP\_INCLXX\_HP}
    \item BERT+PC: \texttt{FTFP\_BERT\_HP+PC}
\end{itemize}
Here, BERT is the Bertini cascade model commonly used in GEANT4, which is used for hadron-nucleus inelastic interactions from 0 to about \SI{6}{\GeV}. The Fritiof parton model (FTF) is used from about \SI{3}{\GeV} to high energies, with a smooth transition in the overlap region. In QGSP-based lists, the quark-gluon string model (QGS) is introduced only at higher energies, starting at about \SI{12}{\GeV} for protons, neutrons, pions, and kaons, while FTF covers the intermediate region. BIC uses a binary cascade treatment based on hadron-nucleus data, while INCL++ is an intranuclear cascade model with a more detailed phase-space treatment. Since selected NC singles are dominated by sub-GeV neutrino interactions and low-energy recoil protons, the relevant SI differences mainly arise from the low-energy cascade treatment, namely BERT, BIC, or INCL++, rather than from the high-energy FTF/QGS string models. Therefore, after their first appearance, we refer to the SI models by their cascade components, namely BERT, BIC, INCL++, and BERT+PC.

\begin{figure}[htbp]
    \centering
    \includegraphics[width=0.95\linewidth]{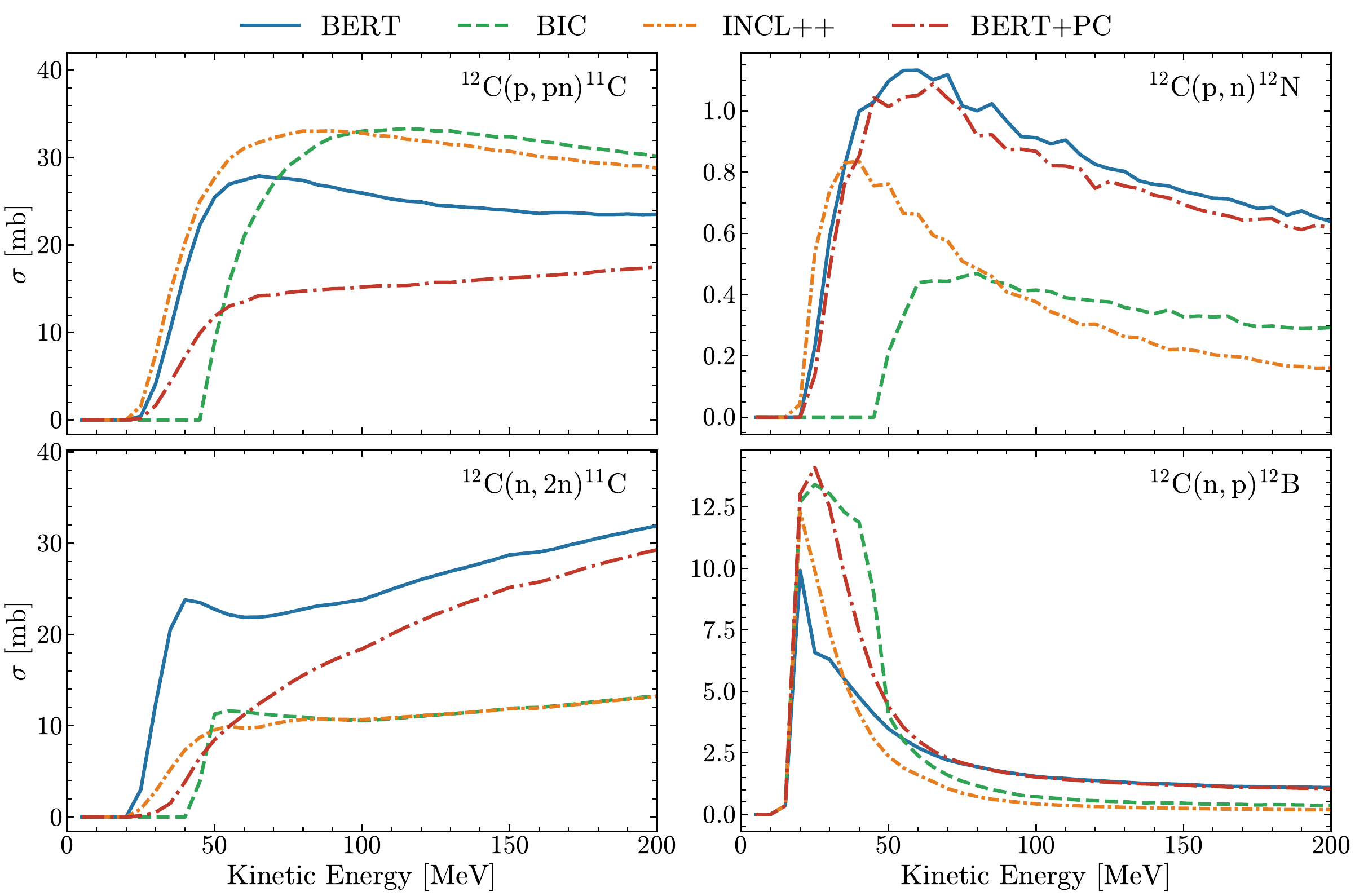}
    \caption{Comparison of GEANT4 hadronic cross sections for four exclusive channels on ${}^{12}\mathrm{C}$ relevant to neutron transport in LS. The cross sections are shown as a function of the incident nucleon kinetic energy. The four curves correspond to the SI treatments listed in Sec.~\ref{sec:processes}: the nominal BERT, BIC, INCL++, and BERT+PC.}
    \label{fig:xsec_summary}
\end{figure}

\begin{figure}[htbp]
    \centering
    \includegraphics[width=0.48\linewidth]{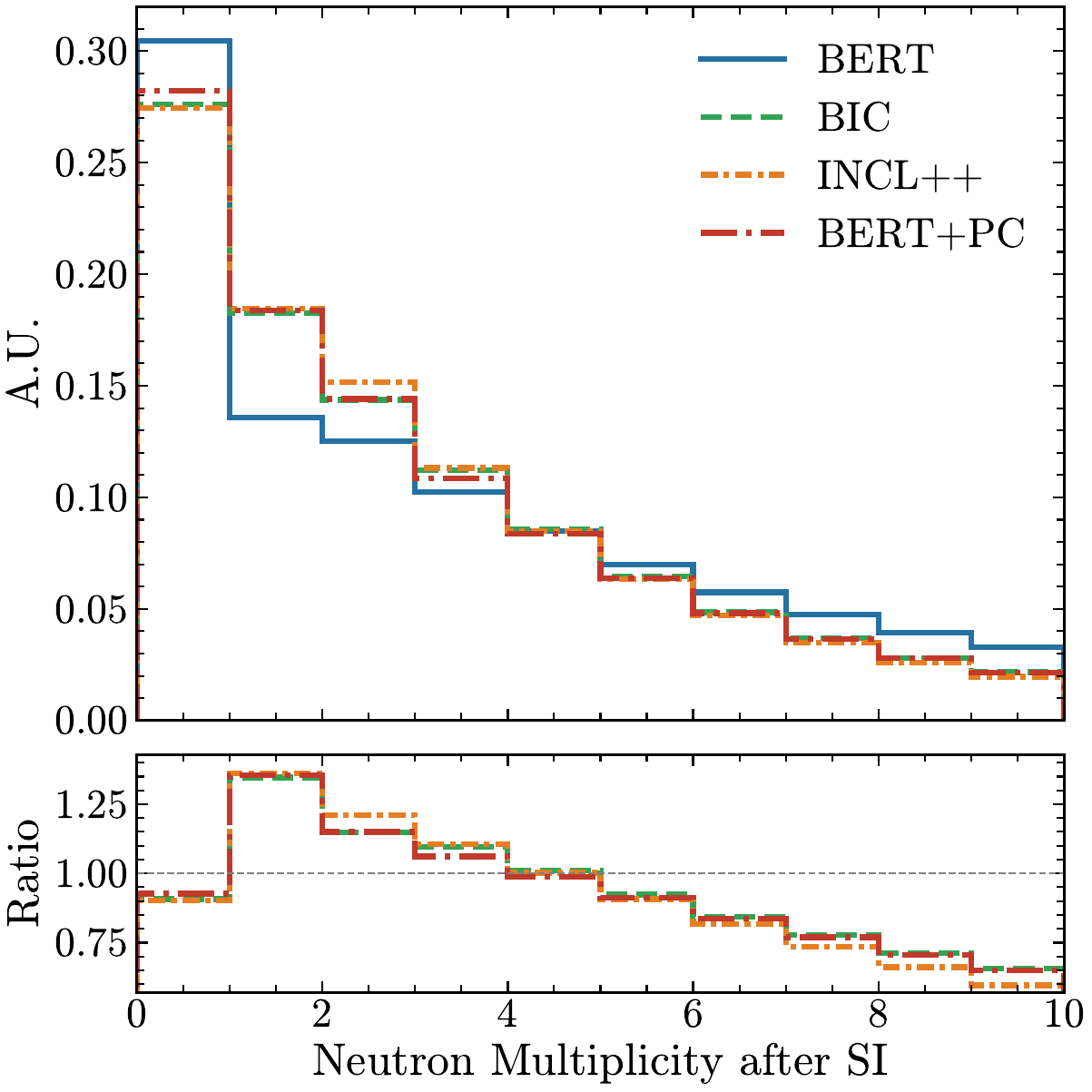}
    \caption{Comparison of captured-neutron multiplicity distributions for different SI models in GEANT4. The distributions are normalized to unit area. The distributions reflect the combined effects of hadronic transport, charge exchange, neutron production, neutron absorption, and post-cascade nuclear de-excitation in LS.}
    \label{fig:si_comp}
\end{figure}

The GEANT4 hadronic cross sections for four exclusive channels on ${}^{12}\mathrm{C}$ relevant to neutron transport in LS are compared in Figure~\ref{fig:xsec_summary}. These channels govern how outgoing protons and neutrons from the primary interaction further interact with carbon nuclei in the scintillator. All four channels can affect the neutron multiplicity: the $(\mathrm{p},\mathrm{p}\mathrm{n})$ and $(\mathrm{n},2\mathrm{n})$ channels produce additional neutrons, while the charge-exchange channels $(\mathrm{p},\mathrm{n})$ and $(\mathrm{n},\mathrm{p})$ change the nucleon content during transport and therefore also modify the final neutron topology. The cross-section differences among the SI models indicate that the microscopic secondary-interaction probabilities are model-dependent. Their combined effect on the captured-neutron multiplicity is discussed below using Figure~\ref{fig:si_comp}.

These SI models can modify the captured-neutron multiplicity distribution, providing a subdominant but non-negligible contribution to the single-signal selection uncertainty. The captured-neutron multiplicity distributions predicted by the different GEANT4 SI models are compared in Figure~\ref{fig:si_comp}. The nominal BERT prediction differs visibly from BIC and INCL++, indicating that the native Bertini cascade and its associated de-excitation treatment produce a different delayed-neutron topology. With the Precompound model enabled, the BERT+PC prediction becomes more consistent with BIC and INCL++, suggesting that part of the BERT difference originates from the post-cascade nuclear de-excitation treatment rather than from the cascade stage alone.

\section{Results}
\label{sec:result}

\subsection{Model Comparisons and Relative Uncertainties}

As mentioned in Sec.~\ref{sec:processes}, we select several models to compare the single-signal predictions across the neutrino-nucleus interaction, de-excitation, and SI. Table~\ref{tab:models} summarizes the model lists used in this study.

\begin{table}
    \centering
    \caption{The model configurations used in this study to compare different processes affecting atmospheric-neutrino singles.}
    \resizebox{\textwidth}{!}{
    \begin{tabular}{c c c c c c c c}
        \toprule
         Model & Generator & Configuration & \makecell{Nuclear\\Model} & $M_{\mathrm{A}}$ [\si{\GeV}] & FSI & De-excitation & SI Model\\
         \midrule
         1 & GENIE & \texttt{G18\_10b\_02\_11b} & LFG & 0.99 & hN & \multirow{6}{*}{TALYS} & \multirow{6}{*}{BERT}\\
         2 & GENIE & \texttt{G18\_10b\_02\_11b} & LFG & 0.99 & hA &  & \\
         3 & GENIE & \texttt{G18\_10d\_02\_11b} & LFG & 0.99 & G4B &  & \\
         4 & GENIE & \texttt{G18\_02b\_02\_11b} & BRRFG & 0.99 & hN &  & \\
         5 & NuWro & default & LFG & 1.03 & Csc &  & \\
         6 & NuWro & default & SF & 1.03 & Csc &  & \\
         \midrule
         7 & \multirow{2}{*}{GENIE} & \multirow{2}{*}{\texttt{G18\_10b\_02\_11b}} & \multirow{2}{*}{LFG} & \multirow{2}{*}{0.99} & \multirow{2}{*}{hN} & No & \multirow{2}{*}{BERT} \\
         8 & & & & & & GEMINI++4$\nu$ & \\
         \midrule
         9 &\multirow{3}{*}{GENIE} & \multirow{3}{*}{\texttt{G18\_10b\_02\_11b}} & \multirow{3}{*}{LFG} & \multirow{3}{*}{0.99} & \multirow{3}{*}{hN}& \multirow{3}{*}{TALYS} & BIC \\
         10 & & & & & & & INCL++ \\
         11 & & & & & & & BERT+PC \\
         \bottomrule
    \end{tabular}
    }
    \label{tab:models}
\end{table}

\begin{figure}[htbp]
    \centering
    \begin{subfigure}{0.95\linewidth}
        \centering
        \includegraphics[width=\linewidth]{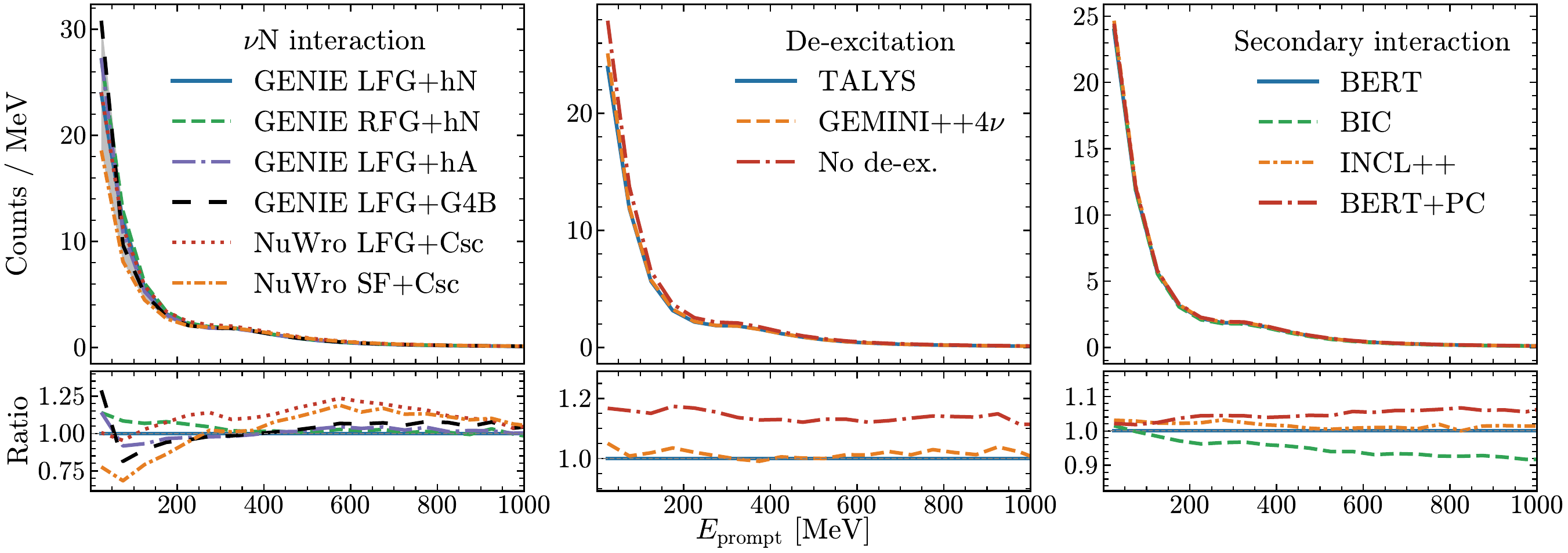}
        \caption{NC.}
    \end{subfigure}
    \begin{subfigure}{0.95\linewidth}
        \centering
        \includegraphics[width=\linewidth]{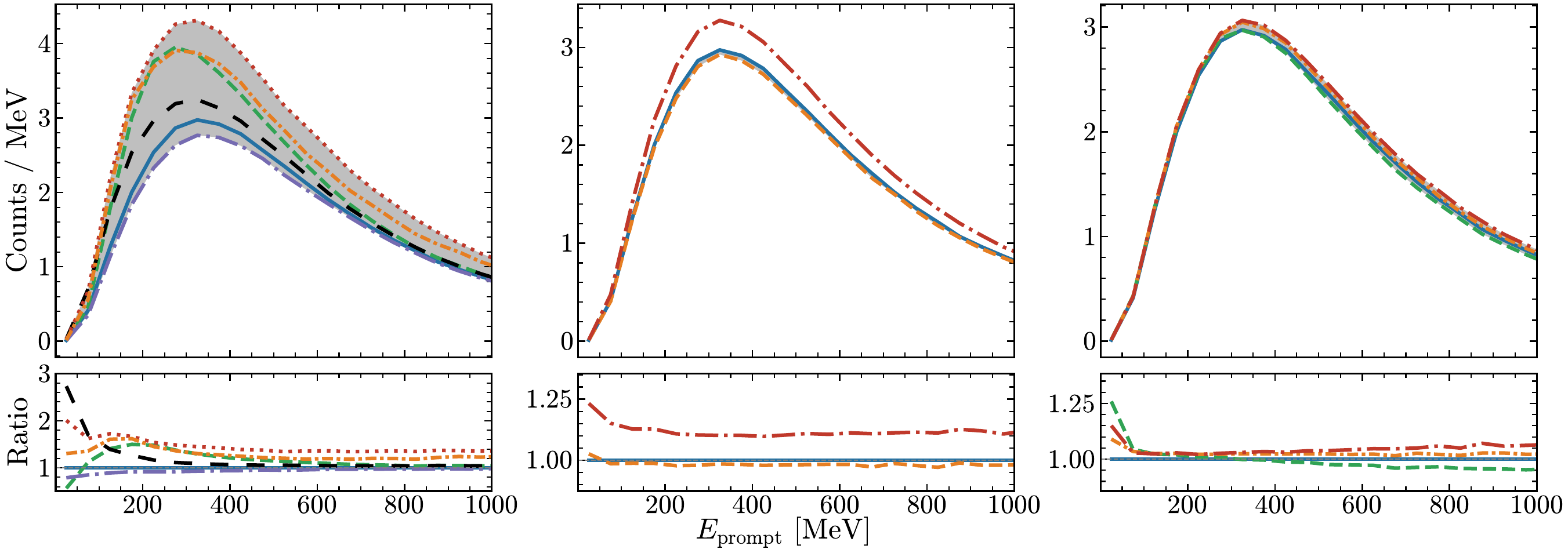}
        \caption{CC.}
    \end{subfigure}
    \caption{Comparison of $E_{\mathrm{prompt}}$ spectra for singles predicted by different model configurations, for an exposure of \SI{200}{\kton\cdot\yr}. The three columns in each panel correspond to the $\nu$N interaction, de-excitation, and secondary interaction stages. The gray bands indicate the envelope of model variations, and the ratio panels show the deviation of each model from the nominal prediction.}
    \label{fig:comp}
\end{figure}

The resulting model dependence of the single-signal $E_{\mathrm{prompt}}$ spectra is summarized in Figure~\ref{fig:comp}, separately for the NC and CC channels. For each channel, the three stages of model dependence---$\nu$N interaction, de-excitation, and secondary interaction---are compared side by side. We find that the largest differences arise from neutrino-nucleus interactions. For NC events, the GENIE hN, hA, and G4B FSI treatments (Models 1--3) are compared. The nuclear-model dependence is examined within each generator by comparing the GENIE LFG and BRRFG variants (Models 1 and 4) and the NuWro LFG and SF variants (Models 5 and 6); the differences here are somewhat more pronounced than those from the FSI variations. Comparing the GENIE and NuWro LFG predictions (Models 1 and 5) reveals the largest difference among the tested configurations. For CC events, nuclear models also have a significant influence, but the primary differences arise from the CCQE models: the Valencia model (Models 1--3) and the Llewellyn-Smith model (Models 4--6). De-excitation is a necessary step for the single-signal prediction, because it determines whether additional neutrons are emitted and whether the event survives the neutron-capture veto. However, the difference between the two realistic de-excitation models, TALYS and GEMINI++4$\nu$, is modest for the selected singles sample. Different SI models, BIC, INCL++, and BERT+PC, result in only a few percent differences relative to the nominal BERT model, and above \SI{100}{\MeV} their spread can become comparable to or larger than the TALYS--GEMINI++4$\nu$ difference.

Based on the comparisons of different models, we estimate the systematic uncertainty in each bin from the envelope of the model predictions, using $(\max-\min)/(\max+\min)$ for the corresponding model class, where $\max$ ($\min$) is the largest (smallest) prediction among the models in that class. For de-excitation, the uncertainty is evaluated only from the difference between TALYS and GEMINI++4$\nu$; the no-de-excitation case is not used in the error budget, because de-excitation is a necessary part of the singles prediction and the no-de-excitation curve is shown only as a diagnostic reference. As an analysis assumption, we treat the flux, primary-interaction, de-excitation, and SI uncertainty sources as independent and calculate the total systematic uncertainty by:
\begin{equation}
    \sigma_{\mathrm{syst}}^2=\sigma_{\mathrm{flux}}^2+\sigma_{\mathrm{\nu N ~ interaction}}^2+\sigma_{\mathrm{deex}}^2+\sigma_{\mathrm{SI}}^2
\end{equation}
where the major contributions to the systematic uncertainty arise from the neutrino flux and neutrino--nucleus interaction models. For the NC sample, the neutrino--nucleus interaction uncertainty is the largest systematic contribution in the low-energy windows. In the $[15,100)$~\si{\MeV} window, it amounts to about $22\%$ of the central value, comparable to the flux uncertainty ($\sim18\%$) and much larger than the statistical uncertainty ($\sim3\%$). In the $[0.7,15)$~\si{\MeV} window, the neutrino--nucleus interaction uncertainty (about $31\%$) similarly exceeds the flux ($\sim23\%$) and statistical ($\sim5\%$) uncertainties by factors of a few. We note that this evaluation is limited to the tested configurations. A reliable estimate of these uncertainties is important for measuring the event rate and energy spectrum of atmospheric-neutrino singles. For example, JUNO started data taking on August 28, 2025~\cite{JUNO:2025fpc}. With its large target mass and excellent energy resolution, JUNO is expected to collect atmospheric-neutrino singles in the coming years. This sample may provide sensitivity to neutrino--nucleus interaction models and their associated uncertainties, which could benefit rare-event searches in liquid-scintillator detectors, such as DSNB and nucleon decay.

\subsection{Predicted Event Rates and Energy Spectra}
\label{subsec:rate_spectra}

\begin{figure}[htbp]
    \centering
    \includegraphics[width=0.48\linewidth]{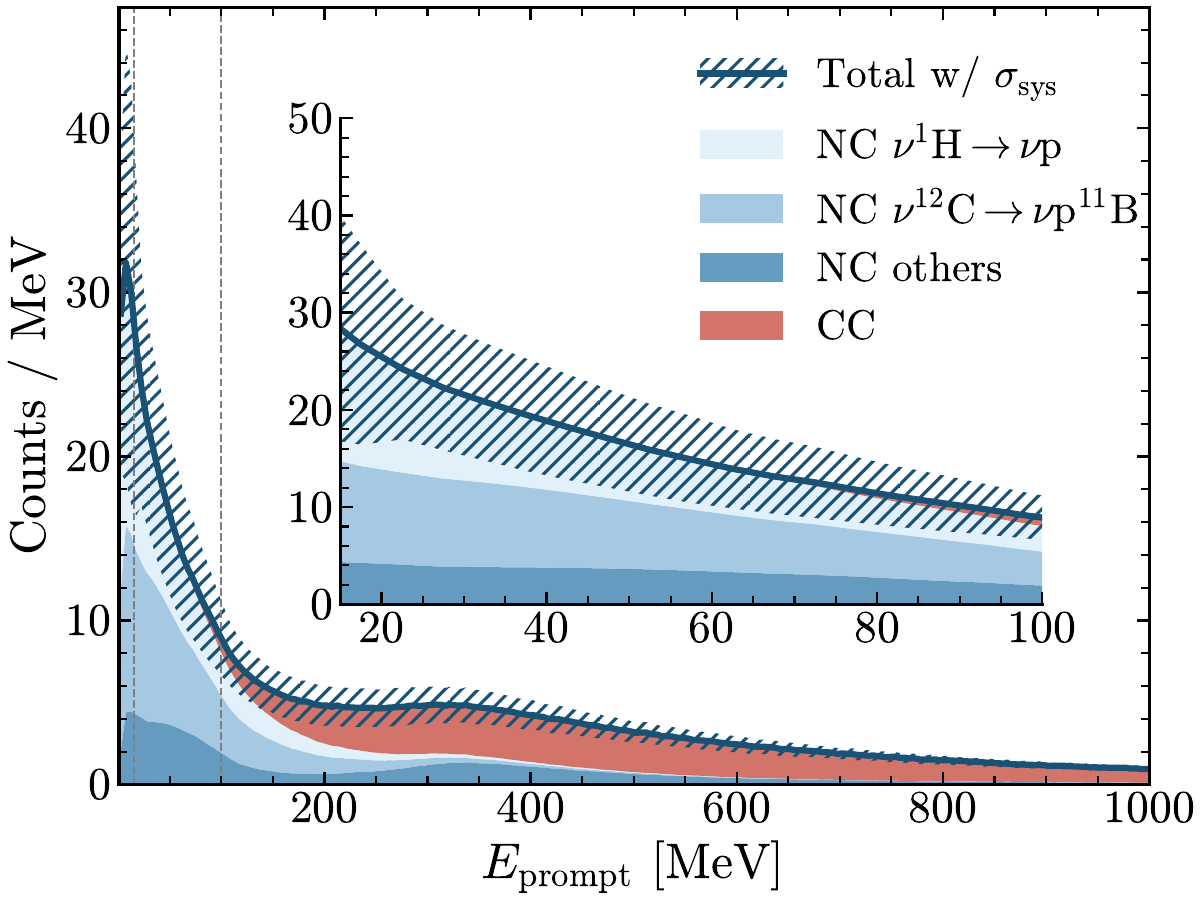}
    \caption{Predicted prompt-energy spectrum of atmospheric-neutrino singles with systematic uncertainties, assuming an exposure of \SI{200}{\kton\cdot\yr}. The solid curve shows the total count, the hatched band indicates the combined systematic uncertainty, and the stacked components show the NC and CC contributions. The inset highlights the low-energy \SIrange{15}{100}{\MeV} region, where the NC component is dominant and the prompt-spectrum composition is especially relevant for background studies.}
    \label{fig:uncertainty}
\end{figure}

Based on the selection criteria in Sec.~\ref{subsec:selection}, we obtain the atmospheric-neutrino singles prediction under the nominal model. The predicted prompt-energy spectrum with systematic uncertainties is presented in Figure~\ref{fig:uncertainty}, together with an inset decomposition of the dominant NC and CC components in the low-energy region. In the $E_{\mathrm{prompt}} < \SI{100}{\MeV}$ range, NC events dominate the selected sample and carry the largest model dependence, while CC events include contributions from both the charged lepton and hadrons and can be identified using pulse-shape discrimination (PSD)~\cite{Cheng:2023zds}. The main backgrounds for atmospheric-neutrino singles are natural radioactivity and cosmogenic isotopes, which are concentrated below $\SI{15}{\MeV}$ and can also be suppressed by PSD. Therefore, LS detectors with good energy resolution could potentially measure the NC singles spectrum below $\SI{100}{\MeV}$ and use it to probe neutrino-nucleus interaction models. At higher energies ($E_{\mathrm{prompt}} \geq \SI{100}{\MeV}$), where PSD is less effective, CC interactions account for the majority of the selected singles, while the NC contribution remains sizable. Because the selection does not distinguish NC and CC interactions on an event-by-event basis, the experimentally accessible observable is the inclusive NC+CC spectrum. Nevertheless, the model-to-model spread in this region is substantially larger for CC than for NC interactions. The sensitivity of the inclusive spectrum to interaction-model variations is therefore dominated by the CC component. Since few other neutrino sources contribute in the sub-GeV range, backgrounds from natural radioactivity can be excluded with an energy cut, and cosmic-ray-muon-related backgrounds can also be vetoed, as discussed in Ref.~\cite{Chauhan:2021fzu}. Further detector-level studies are needed to quantify these prospects.

Atmospheric-neutrino singles can also constitute backgrounds to other prompt-only signatures in LS detectors. For example, searches for solar ${}^{8}\mathrm{B}$ neutrinos based on $\nu$-e elastic scattering receive a background contribution from atmospheric-neutrino singles with $E_{\mathrm{prompt}}\in[0.7,15)$~\si{\MeV}. New-physics scenarios, such as dark-matter annihilation into sterile neutrinos and boosted dark matter, can also be affected by atmospheric-neutrino singles in the $E_{\mathrm{prompt}}\in[15,100)$~\si{\MeV} range. At higher energies, many nucleon-decay channels, in which prompt products such as $e^\pm$ and $\pi^0$ deposit their energy immediately, can receive backgrounds from atmospheric-neutrino singles at several hundred MeV. The corresponding counts of NC and CC singles in these prompt-energy ranges are summarized in Table~\ref{tab:sys_breakdown}.

Table~\ref{tab:sys_breakdown} lists the individual systematic-uncertainty sources for each prompt-energy window.

\begin{table}[htbp]
    \centering
    \caption{Individual systematic-uncertainty sources for the predicted single-signal event counts in each prompt-energy window (\SI{10}{\yr} exposure in a \SI{20}{\kton} LS detector).}
    \begin{tabular}{l c c c c c c}
        \toprule
        & \multicolumn{3}{c}{NC} & \multicolumn{3}{c}{CC} \\
        \cmidrule(lr){2-4}\cmidrule(lr){5-7}
        $E_{\mathrm{prompt}}$ (\si{\MeV}) & $[0.7,15)$ & $[15,100)$ & $[100,1000)$ & $[0.7,15)$ & $[15,100)$ & $[100,1000)$ \\
        \midrule
        Event count & 453 & 1337 & 1066 & $\lesssim 0.01$ & 21 & 1750 \\
        $\sigma_{\mathrm{flux}}$ & 103 & 245 & 114 & $\lesssim 0.01$ & 7 & 271 \\
        $\sigma_{\nu\mathrm{N}}$ & 142 & 289 & 76 & $\lesssim 0.01$ & 7 & 355 \\
        $\sigma_{\mathrm{deex}}$ & 9 & 23 & 8 & $\lesssim 0.01$ & 0 & 17 \\
        $\sigma_{\mathrm{SI}}$ & 6 & 20 & 39 & $\lesssim 0.01$ & 1 & 45 \\
        $\sigma_{\mathrm{syst}}$ & 175 & 380 & 143 & $\lesssim 0.01$ & 10 & 449 \\
        \bottomrule
    \end{tabular}
    \label{tab:sys_breakdown}
\end{table}

\subsection{Contribution from Low-Energy Atmospheric Neutrinos}

We have so far considered atmospheric-neutrino singles from neutrinos with $E_\nu\ge\SI{100}{\MeV}$. Here, we additionally consider the contribution from neutrinos with $E_\nu<\SI{100}{\MeV}$, restricting the simulation to elastic scattering on free protons. The sub-\SI{100}{\MeV} atmospheric-neutrino fluxes are provided by the Honda group without accounting for the mountain overburden at the JUNO site. This assumption may have a significant impact on the flux in this energy range~\cite{Guo:2018sno,Cheng:2026cqx} and should therefore be kept in mind when interpreting this contribution.

At these low neutrino energies, the nuclear-model treatment implemented in the event generators is not reliable for the very low-energy recoils relevant here. We therefore calculate the kinetic energy of the final-state proton from $\nu\mathrm{p}$ elastic scattering (PES) using Eq.~\ref{ahren} and use this to model the detector response. Because of the small momentum transfer in $\nu\mathrm{p}$ elastic scattering, the resulting recoil protons have very low kinetic energies. We find that, for an exposure of \SI{200}{\kton\cdot\yr}, approximately 78 such events contribute to the singles sample with $E_{\mathrm{prompt}}<\SI{15}{\MeV}$. This region is dominated by natural-radioactivity backgrounds, and the present study focuses on the \SIrange{15}{100}{\MeV} prompt-energy window. We therefore do not further evaluate the systematic uncertainties associated with this low-energy contribution.

\begin{figure}[htbp]
    \centering
    \includegraphics[width=0.48\linewidth]{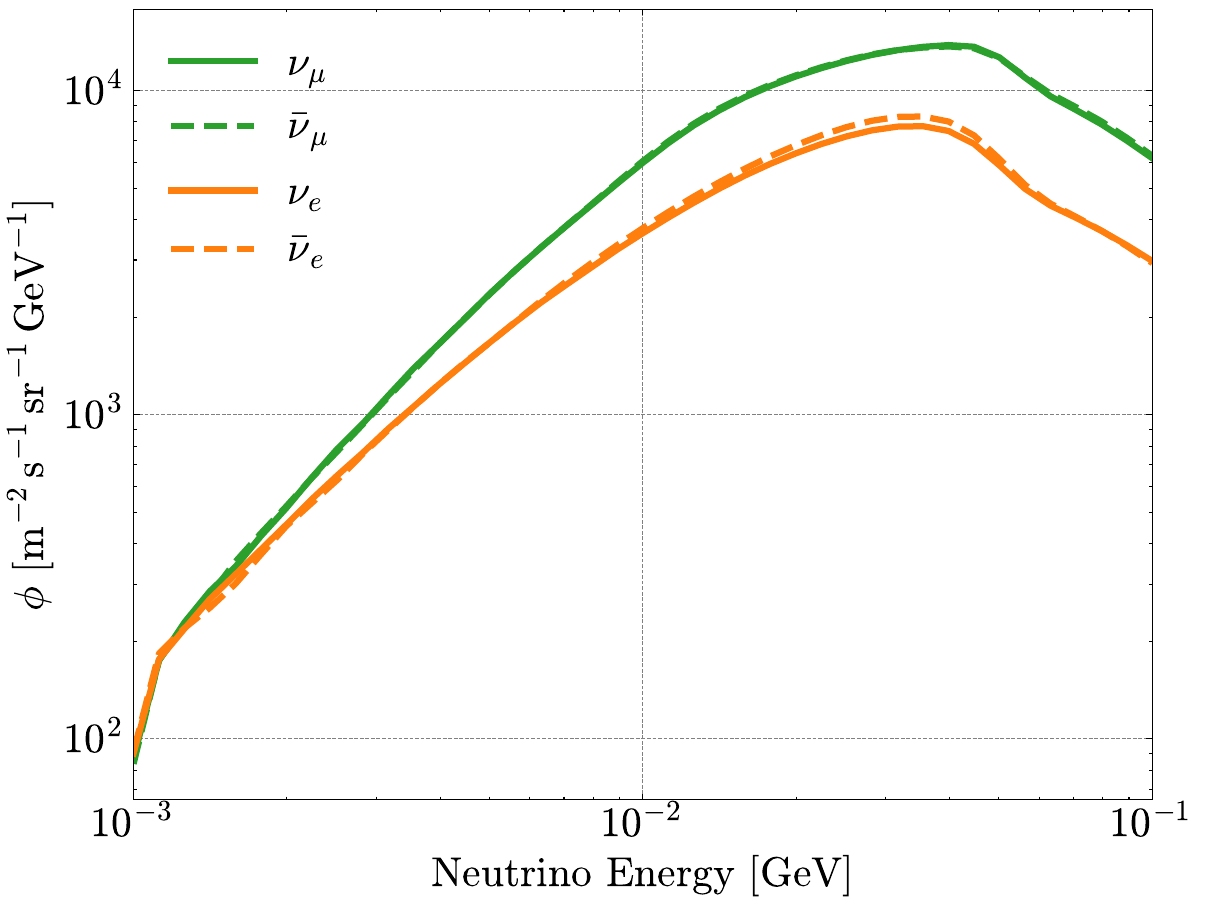}
    \includegraphics[width=0.48\linewidth]{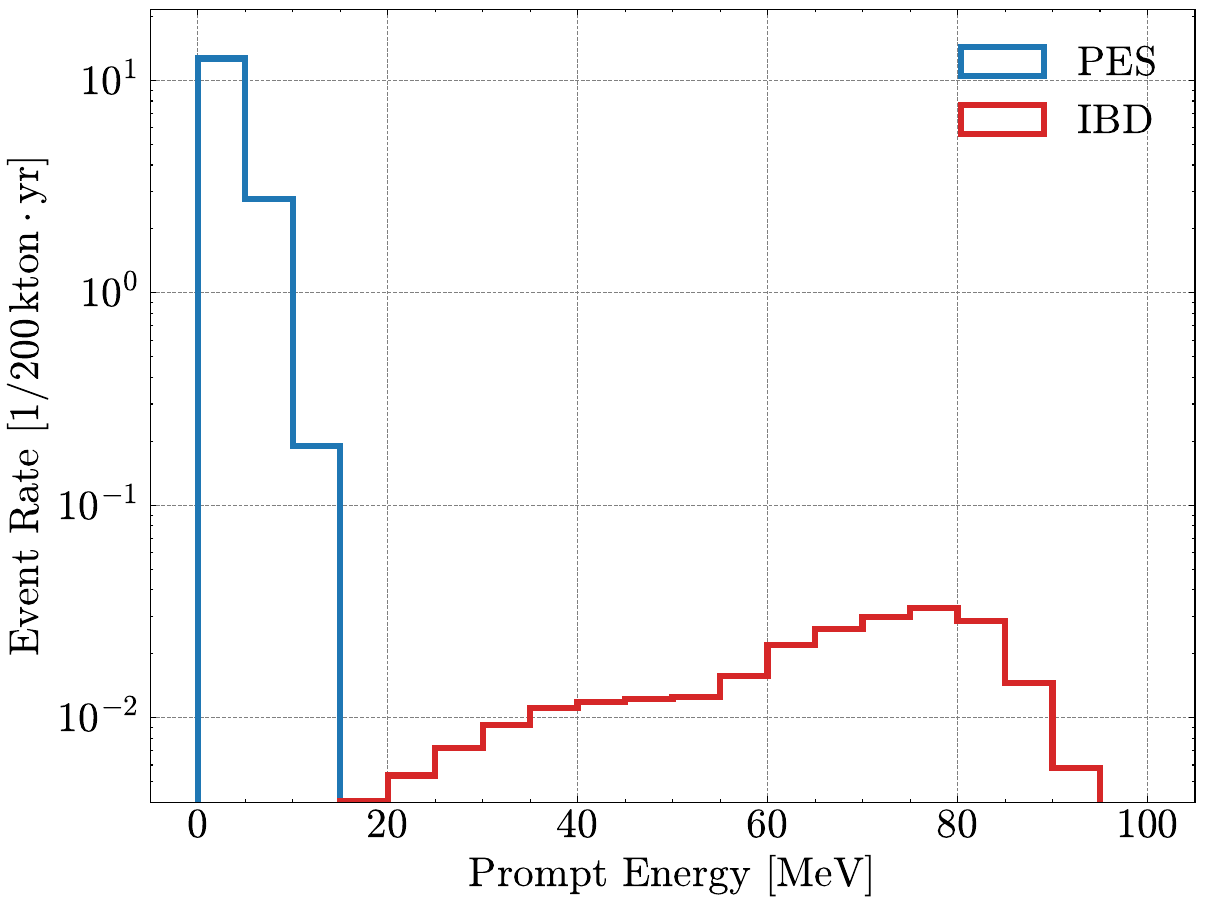}
    \caption{Low-energy atmospheric-neutrino fluxes and the resulting energy spectra of singles from various processes after selection. PES is the dominant contributor to singles below $\SI{15}{\MeV}$.}
    \label{fig:le}
\end{figure}

We also simulate IBD events using the cross section of Ref.~\cite{Strumia:2003zx}. As shown in Figure~\ref{fig:le}, the IBD contribution to the selected singles sample is negligible.

Low-energy neutrinos can also interact with ${}^{12}\mathrm{C}$, although a detailed treatment is beyond the scope of this work. Reference~\cite{Suliga:2023pve} investigates the exclusive NC excitation of ${}^{12}\mathrm{C}$ as a signal of sub-\SI{100}{\MeV} atmospheric neutrinos, identified through the ensuing \SI{15.11}{\MeV} $\gamma$ ray. The leading reducible background to this $\gamma$-line search arises from inclusive atmospheric-neutrino interactions that produce proton-recoil singles of the type studied in this work.

\section{Implications and Discussion}
\label{sec:discussion}

\subsection{Backgrounds for Rare Event Searches}

\noindent\textbf{Indirect Dark Matter Detection.}
Atmospheric-neutrino singles are also relevant as backgrounds for indirect dark matter searches in large LS detectors, especially for prompt-only signatures in the tens-of-MeV to sub-GeV visible-energy range. The study in Ref.~\cite{Chauhan:2021fzu} provides an early exploration of this idea and demonstrates that such prompt-only event selections can be used to probe dark-matter-related scenarios. In this context, our results suggest that the accuracy of the atmospheric singles prediction is important not only for rate estimates but also for the sensitivity of these searches, since the underlying model dependence can propagate directly into the expected background level.

\noindent\textbf{$0\nu\beta\beta$ Search.} 
Large LS detectors are promising for neutrinoless double-beta decay ($0\nu\beta\beta$) searches thanks to their capability to load large quantities of isotopes. For example, JUNO plans an upgrade specifically targeting $0\nu\beta\beta$ detection~\cite{juno_0vbb}. In the region of interest (ROI) around \SI{2.5}{\MeV} for ${}^{136}\mathrm{Xe}$ and ${}^{130}\mathrm{Te}$, the NC-induced background is estimated in this work to be about $3\times10^{-7}~\si{[keV \cdot (ton~LS) \cdot yr]^{-1}}$. This background can be significantly suppressed using PSD to differentiate electron signals (from $0\nu\beta\beta$) from those produced by NC interactions involving proton recoils. Given this effective background reduction via PSD, NC events are expected to be a negligible background for future high-sensitivity $0\nu\beta\beta$ searches in LS detectors such as JUNO.

\subsection{Calibration Samples} 
The sample of singles below \SI{100}{\MeV} is enriched in proton-recoil events, as shown in Figure~\ref{fig:uncertainty}. This sample serves as a crucial control sample for validating the simulated timing distributions of recoil protons in the tens-of-MeV energy range. Such validation is essential because NC-induced IBD-like events constitute the dominant background for the DSNB search. Since the DSNB signal is a correlated positron--neutron pair, distinguishing it from this background relies heavily on accurately modeling the PMT hit time distribution of the recoil protons, which must be separated from the correlated positron signals.

Particularly useful channels are $\nu+{}^{1}\mathrm{H}\rightarrow\nu+\mathrm{p}$ and $\nu+{}^{12}\mathrm{C}\rightarrow\nu+\mathrm{p}+{}^{11}\mathrm{B}$, where a single recoil proton carries most of the visible energy after quenching. These channels therefore provide relatively clean samples for validating proton-recoil timing and pulse-shape models in LS. The same sample is also sensitive to the scintillator quenching response and, when combined with dedicated calibration data, can provide a complementary check of the Birks-law parameters relating proton kinetic energy to visible energy. Other NC channels below $\SI{100}{\MeV}$ can contain multiple protons, such as $\nu+{}^{12}\mathrm{C}\rightarrow\nu+2\mathrm{p}+{}^{10}\mathrm{Be}$, $\nu+{}^{12}\mathrm{C}\rightarrow\nu+3\mathrm{p}+{}^{9}\mathrm{Li}$, and $\nu+{}^{12}\mathrm{C}\rightarrow\nu+4\mathrm{p}+{}^{8}\mathrm{He}$. These multi-proton topologies have more complicated quenching, timing, and vertex properties, but they remain useful for cross-checking the detector response to hadronic prompt signals. NC channels with pion production can be ignored in this low-energy control sample: assuming no energy leakage, events with $\pi^\pm$ or $\pi^0$ rarely contribute to $E_{\mathrm{prompt}}<\SI{100}{\MeV}$.

\subsection{Neutrino-Nucleus Interaction Model Constraints}

\begin{figure}[htbp]
    \centering
    \includegraphics[width=0.48\linewidth]{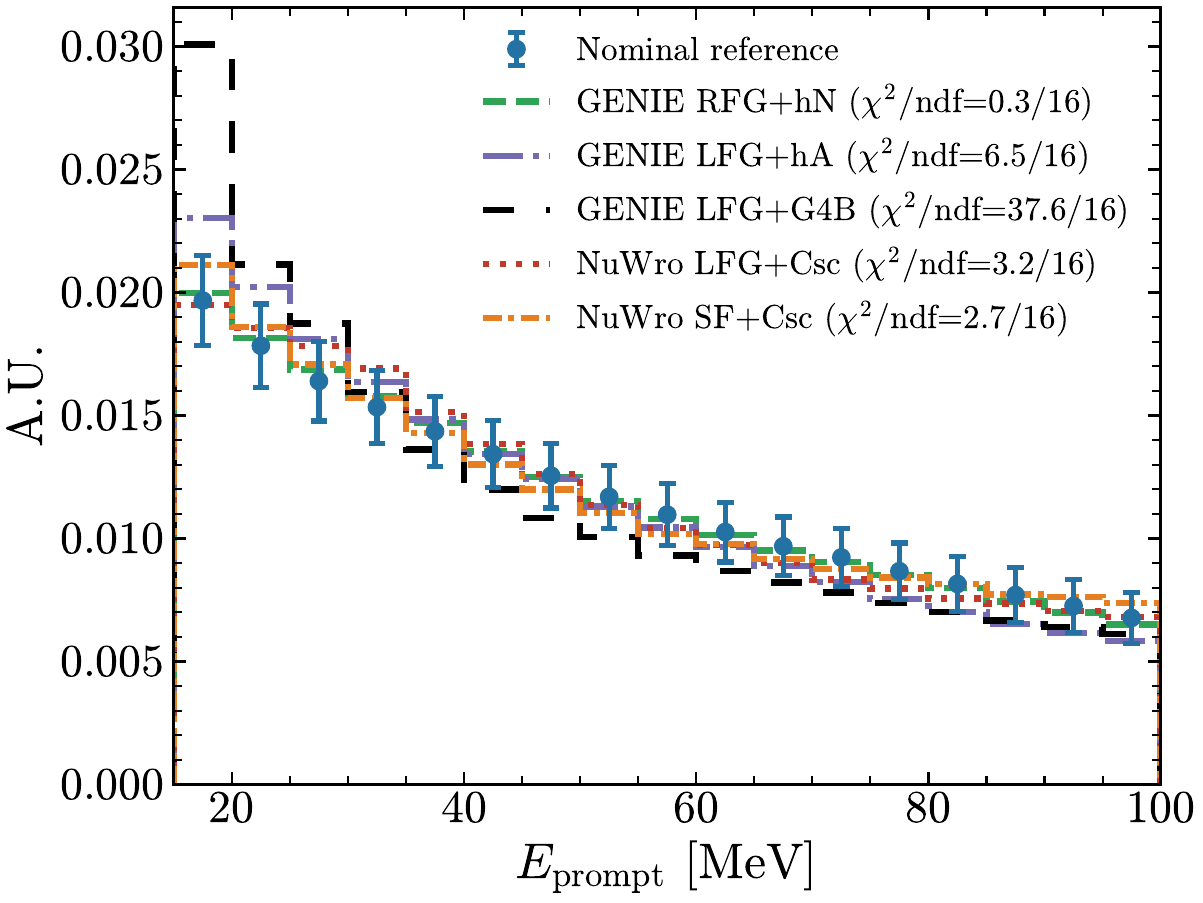}
    \caption{Shape-only comparison of the selected atmospheric-neutrino singles prompt-energy spectrum in the \SIrange{15}{100}{\MeV} region. The nominal GENIE LFG prediction is used as the reference, while alternative neutrino-interaction models are normalized to the same integral in this energy window. The normalization is fixed independently in this window, so the comparison tests the shape rather than the total rate. The error bars include statistical, flux-shape, de-excitation, and secondary-interaction uncertainties.}
    \label{fig:shape_chi2_15_100}
\end{figure}

The selected singles spectrum can also be used to test neutrino-nucleus interaction models. The $E_{\mathrm{prompt}}\in[15,100)$~\si{\MeV} region is particularly useful because the sample is NC-dominated and less affected by natural radioactivity than the sub-\SI{15}{\MeV} region. To illustrate this sensitivity, Figure~\ref{fig:shape_chi2_15_100} shows a shape-only comparison in this energy window. The nominal GENIE LFG prediction is used as the reference, and the alternative interaction models are normalized to the same integral before comparison. This removes the overall normalization difference and emphasizes the prompt-energy shape induced by the primary interaction model.

To quantify the difference in shape, we compute the covariance-weighted $\chi^2$ between the normalized shape of each alternative model and the nominal prediction:
\begin{equation}
\label{eq:chi2}
    \chi^2 = \left(s_{\mathrm{model}} - s_{\mathrm{nominal}}\right)^{T}\,C^{-1}\,\left(s_{\mathrm{model}} - s_{\mathrm{nominal}}\right),
\end{equation}
where $s$ denotes the prompt-energy shape vector in the analysis window, normalized to unit integral. The covariance matrix $C$ includes the statistical covariance, evaluated using multinomial statistics for an exposure of \SI{200}{\kton\cdot\yr}, and the systematic covariance contributions from flux-shape, de-excitation, and secondary-interaction uncertainties. For the flux-shape source, the covariance is constructed from the $1\sigma$ shape shift. For the de-excitation and secondary-interaction sources, the covariance is constructed from the bin-by-bin shape differences between the alternative and nominal models, with $C^{\mathrm{syst}}_{ij}=\Delta s_i\,\Delta s_j$ for each variation. Because the shape vectors are normalized to unit integral, the total covariance matrix is singular along the normalization direction. We therefore use its pseudoinverse, $C^{+}$, in Eq.~\ref{eq:chi2}. The number of independent degrees of freedom is $N_{\mathrm{bin}}-1$, reflecting the normalization constraint. The comparison is thus sensitive only to differences in shape. The nominal prediction is used as the reference expectation without statistical fluctuations. A small $\chi^2/\mathrm{ndf}$ indicates that the difference between the alternative and nominal shapes is small relative to the combined statistical and systematic uncertainties.

In this normalized shape-only comparison, the GENIE LFG+G4B configuration gives the largest deviation from the nominal prediction, with $\chi^2/\mathrm{ndf}=2.35$. This indicates that the selected \SIrange{15}{100}{\MeV} singles shape is particularly sensitive to the intranuclear FSI treatment implemented in the generator. In contrast, the GENIE RFG+hN, NuWro LFG+Csc, and NuWro SF+Csc predictions give much smaller deviations, with $\chi^2/\mathrm{ndf}=0.02$, $0.20$, and $0.17$, respectively. These results suggest that the nuclear-model differences tested here are difficult to distinguish using the normalized prompt-energy shape alone. The GENIE LFG+hA model gives an intermediate difference, $\chi^2/\mathrm{ndf}=0.41$, further supporting the role of FSI in shaping this energy region. These results do not contradict the larger rate-level model dependence discussed in Sec.~\ref{sec:result}; rather, they indicate that a shape-only measurement of atmospheric-neutrino singles could provide complementary, data-driven sensitivity to $\nu\mathrm{C}$ interaction modeling, especially to the FSI treatment, once detector effects and backgrounds are incorporated into a dedicated sensitivity study.

\section{Summary}
\label{sec:sum}

Atmospheric-neutrino interactions in large LS detectors are strongly model-dependent, especially for NC processes. Unlike low-energy neutrino sources such as reactor neutrinos, atmospheric neutrinos cover a wide energy range and allow the measurement of prompt-only events, namely singles, without requiring coincidence signals. In this paper, we systematically study the effects of different models on atmospheric-neutrino singles and predict the $E_{\mathrm{prompt}}$ spectrum together with the associated systematic uncertainties.

We separate the modeling into three stages: the neutrino-nucleus interaction, residual-nucleus de-excitation, and secondary interactions in the scintillator. Among the tested model configurations, the nuclear effects in $\nu\mathrm{C}$ interactions produce the primary differences in both NC and CC events, while the variations between the Llewellyn-Smith and Valencia CCQE models are also notable. De-excitation is a necessary ingredient for singles because the selection is highly sensitive to neutron multiplicity, but realistic de-excitation models lead to similar $E_{\mathrm{prompt}}$ spectra. Secondary-interaction models are likewise subdominant, since the final-state particles in singles are mainly recoil protons (and $e^\pm$ in CC events), which lose energy primarily through ionization rather than through processes that change the neutron multiplicity.

The results presented in this paper have broad applicability. Future large LS detectors, such as JUNO, may be able to measure atmospheric-neutrino singles. For these detectors, predictions for neutrino scattering on free protons show a relatively small spread among the tested configurations, while the $\nu\mathrm{C}$ predictions carry larger model-driven uncertainties, indicating potential sensitivity to $\nu\mathrm{C}$ interaction models, particularly for NC interactions. The selected low-energy singles sample could also serve as a control sample for validating proton-recoil timing, pulse-shape, and quenching responses in LS. In addition, atmospheric-neutrino singles will contribute to the backgrounds in studies that focus on prompt signal events in LS, such as solar ${}^{8}\mathrm{B}$ neutrino searches and new physics scenarios, including dark matter annihilation to sterile neutrinos, boosted dark matter, and nucleon decays. The predictions in this paper, together with future data on atmospheric-neutrino singles from large LS detectors, could provide new insights into atmospheric-neutrino interactions and support studies of more complex event topologies.
\section*{Acknowledgments}

The authors thank Nahid Bhuiyan for helpful discussions on neutron capture in GEANT4, and Qiyu Yan, Xianguo Lu, and Costas Andreopoulos for their help with the GENIE and NuWro generators. The authors also thank Qiyu Yan and Hongyue Duyang for their careful reading of the manuscript and their valuable comments and suggestions.

This work was supported in part by the National Key R\&D Program of China under Grant No.\ 2024YFE0110500, the National Natural Science Foundation of China under Grant Nos.\ 12375098 and 12125506, and the CAS Project for Young Scientists in Basic Research under Grant No.\ YSBR-099.

\appendix
\section{Atmospheric-Neutrino IBD-Like Events}

In addition to the single-signal topology, atmospheric-neutrino interactions can produce IBD-like events with one prompt signal and one selected neutron-capture signal. This prompt-delayed topology is relevant for reactor-neutrino and DSNB searches in LS detectors. We use the same selection as for singles, except that the selected captured-neutron multiplicity is required to be one instead of zero.

\begin{figure}[htbp]
    \centering
    \includegraphics[width=0.48\linewidth]{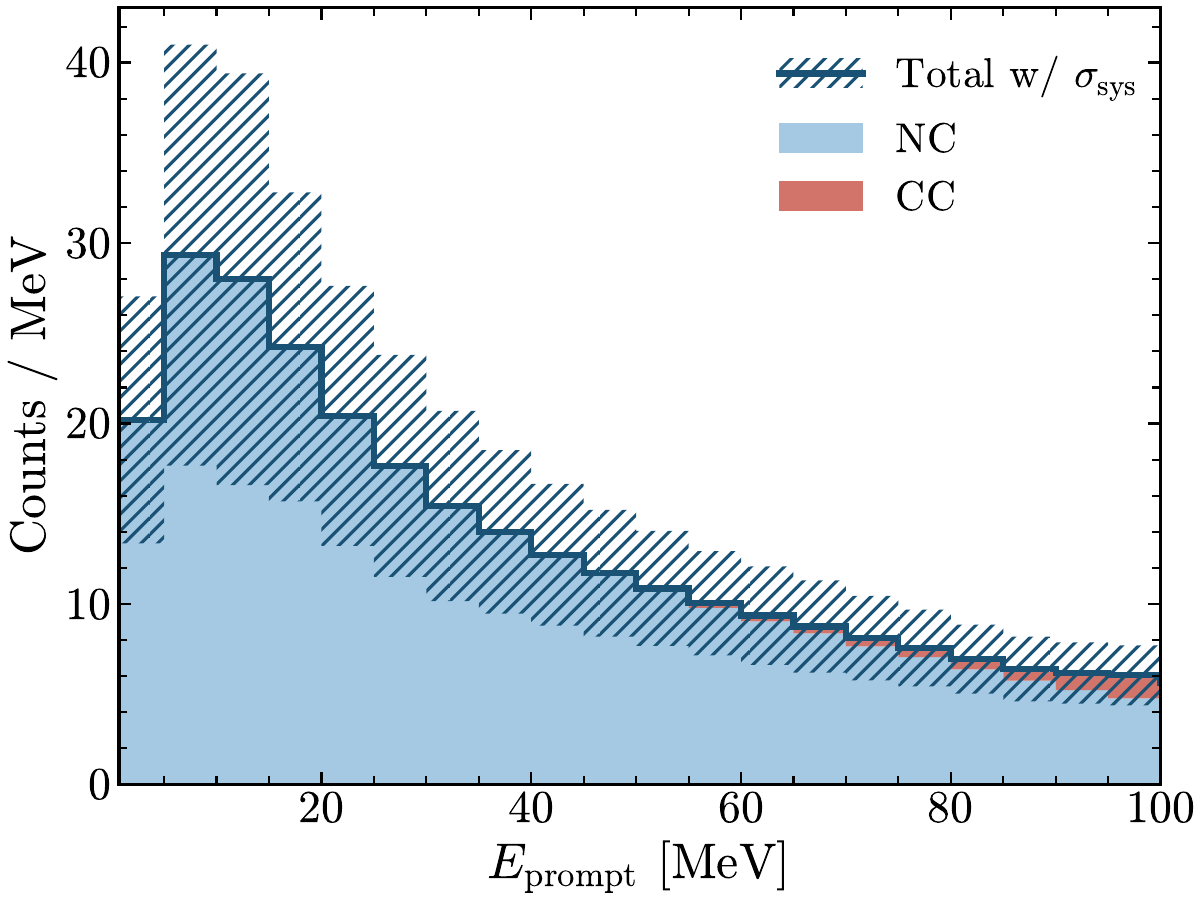}
    \caption{Predicted prompt-energy spectrum of atmospheric-neutrino IBD-like events with one selected neutron-capture signal in the \SIrange{0.7}{100}{\MeV} region, assuming an exposure of \SI{200}{\kton\cdot\yr}. The uncertainty band shows the systematic uncertainty.}
    \label{fig:uncertainty_n1}
\end{figure}

The prompt-energy threshold, fully contained requirement, neutron-capture time and energy windows, and Michel-electron veto are unchanged. The systematic uncertainty is evaluated with the same model-envelope method as in the main text. Figure~\ref{fig:uncertainty_n1} shows the predicted prompt-energy spectrum for an exposure of \SI{200}{\kton\cdot\yr}.

The integrated rates are summarized in Table~\ref{tab:ibd_like_rate}. The \SIrange{0.7}{12}{\MeV} region is relevant for reactor-neutrino measurements, while \SIrange{12}{30}{\MeV} corresponds to the DSNB search window.

\begin{table}
    \centering
    \caption{Predicted rates of atmospheric-neutrino IBD-like events. The uncertainties are systematic.}
    \begin{tabular}{c c c c}
        \toprule
        $E_{\mathrm{prompt}}$ [\si{\MeV}] & \numrange{0.7}{12} & \numrange{12}{30} & \numrange{0.7}{100} \\
        \midrule
        Rate [\si{\per\kton\per\yr}] & $1.52 \pm 0.56$ & $1.97 \pm 0.69$ & $6.85 \pm 2.04$ \\
        \bottomrule
    \end{tabular}
    \label{tab:ibd_like_rate}
\end{table}

The rates are close to the comprehensive atmospheric NC background prediction in Ref.~\cite{Cheng:2024uyj}, but small differences remain. These differences are expected because the present calculation uses updated generators and GEANT4 simulation configurations.

\bibliographystyle{unsrt}
\bibliography{ref}

\end{document}